\documentstyle[amsfonts,amssymb,amsfonts,amssymb,graphicx,mathrsfs,color,ulem]{article}

\newcommand{\R}{{\mathbb R}}
\newcommand{\C}{{\mathbb C}}

\newcommand{\dsp}{\displaystyle}

\newcommand{\ol}{\overline}
\newcommand{\ndt}{\noindent}
\newcommand{\qed}{\hfill $\square$}

\newtheorem{theorem}{Theorem}[section]
\newtheorem{lemma}[theorem]{Lemma}

\newtheorem{corollary}[theorem]{Corollary}
\newtheorem{remark}[theorem]{Remark}

\title{\bf 
Global in time   self-interacting Dirac fields in the de~Sitter space}

\author{{\bf Karen Yagdjian} }

\begin{document}

\date{}
\maketitle
\thispagestyle{empty}
\vspace{-0.3cm}

\begin{center}
{\it School of Mathematical and Statistical Sciences,\\
University of Texas RGV,\\
1201 W.~University Drive,  
Edinburg, TX 78539,
USA }
\end{center}
\medskip

\addtocounter{section}{-1}
\renewcommand{\theequation}{\thesection.\arabic{equation}}
\setcounter{equation}{0}
\pagenumbering{arabic}
\setcounter{page}{1}
\thispagestyle{empty}

\hspace{2cm}\begin{abstract}
\begin{small}
In this paper the semilinear equation of the spin-$\frac{1}{2}$  fields in the de~Sitter  space is investigated.
We prove the existence  of the global in time  small data solution in the expanding de~Sitter universe. Then, under the Lochak-Majorana condition, we prove  the    existence  of the global in time solution with large data.  The sufficient conditions for the solutions to blow  up in finite time are given for large data in   the expanding and contracting  de~Sitter spacetimes. The influence of the Hubble constant on the lifespan  is estimated.    
\medskip

\end{small}
\end{abstract}

\setcounter{equation}{0}
\renewcommand{\theequation}{\thesection.\arabic{equation}}

\section{Introduction}

In this article we study the  solvability and solutions of the semilinear Dirac  equation in the de~Sitter space, which   is a member of the curved spacetimes of the Friedmann-Lema\^itre-Robertson-Walker (FLRW) models of cosmology (see, e.g., \cite{Gron-Hervik,Ohanian-Ruffini}).  More precisely, we derive   conditions on the mass term, Hubble constant, nonlinear term, and initial function, which guarantee either global in time existence  or an estimate of the  lifespan  of the solution  to the Dirac equation in the expanding or contracting de~Sitter universe.

Although the linear Dirac equation in the curved spacetime was known for a long time (see, e.g., \cite{Birrell,Fock,Parker,Schrodinger,Thaller}) and  was well-investigated, and the semilinear 
Dirac equation in the Minkowski space was  the focus of many publications (see, e.g., \cite{Bachelot_1988,Bachelot,Candy,Piero,Shuji} and references therein), the semilinear Dirac equation in curved spacetime is, to the best of our knowledge,   lacking any study.  The latest astronomical  discovery that the expansion of the universe is accelerating  underscores the need to consider the  equations of the quantum field theory  in the curved spacetimes, especially  in the inflationary theories of the early universe. We are motivated by the  importance of the   spin-$\frac{1}{2}$ particles in the expanding 
universe.

The metric tensor of the spatially flat  de~Sitter space 
has the scale factor   $a(t)=e^{Ht}  $ (see, e.g., \cite{Moller}), which    models  the expanding or contracting universe if $H>0$ or $H<0$, respectively. The curvature of this space is $-12H^2$. The Dirac equation in the de~Sitter space is (see, e.g., \cite{Barut-D})
\[
  \dsp 
\left(  i {\gamma }^0    \partial_0   +i e^{-Ht}{\gamma }^1  \partial_1+i  e^{-Ht}{\gamma }^2 \partial_2+i e^{-Ht}{\gamma }^3   \partial_3 +i \frac{3}{2}    H {\gamma }^0     -m{\mathbb I}_4 \right)\psi=f \,,
\]
where 
the contravariant gamma matrices are   (see,  e.g., \cite[p. 61, Sec.6.2]{B-Sh}) 
\begin{eqnarray*}
&  &
 \gamma ^0= \left (
   \begin{array}{ccccc}
   {\mathbb I}_2& 0   \\
   0& -{\mathbb I}_2   \\ 
   \end{array}
   \right),\quad 
\gamma ^k= \left (
   \begin{array}{ccccc}
  {\mathbb O}_2& \sigma ^k   \\
  -\sigma ^k &  {\mathbb O}_2  \\  
   \end{array}
   \right),\quad k=1,2,3,\\
&  &
 \gamma ^5:=-i\gamma ^0\gamma ^1\gamma ^2\gamma ^3=\left(
\begin{array}{cccc}
{\mathbb O}_2 & -{\mathbb I}_2 \\
 -{\mathbb I}_2 & {\mathbb O}_2 \\
\end{array}
\right) \,\, \mbox{chirality matrix} \,.  
 \end{eqnarray*}
Here $\partial_0 = \frac{\partial}{\partial t} $, $\partial_k = \frac{\partial}{\partial x_k} $, $k=1,2,3$, while  $\sigma ^1 $, $\sigma ^2 $, and $\sigma ^3 $   are the Pauli matrices 
\begin{eqnarray*}
&  &
\sigma ^1= \left (
   \begin{array}{ccccc}
  0& 1   \\
  1& 0  \\  
   \end{array}
   \right), \quad
\sigma ^2= \left (
   \begin{array}{ccccc}
  0& -i   \\
  i& 0  \\  
   \end{array}
   \right),\quad
\sigma ^3= \left (
   \begin{array}{ccccc}
  1& 0   \\
  0&-1 \\  
   \end{array}
   \right)\,,
\end{eqnarray*}
and  ${\mathbb I}_n $, ${\mathbb O}_n $ denote the $n\times n$ identity and zero matrices, respectively. 

Consider the Dirac equation
\begin{equation}
\label{LDE}
  \dsp 
\left(  i {\gamma }^0    \partial_0   +i e^{-Ht}\sum_{\ell=1,2,3}{\gamma }^\ell  \partial_\ell +i \frac{3}{2}    H {\gamma }^0     -m{\mathbb I}_4 +{\gamma }^0V(x,t) \right)\psi=0  \,,
\end{equation}
where the matrix-valued potential $  V(x,t): {\mathbb R}^4 \longrightarrow M_4({\mathbb C}) $ has the following structure
\[
 V(x,t) = \sum_{\ell=1,2,3}A_\ell(x,t) \alpha ^j +A_0(x,t){\gamma }^0 +V_0 (x,t) \,.
\]
In particular, it can be generated by the electromagnetic potential $(A_0(x,t),A_\ell(x,t)) $  (the magnetic potential $\vec A$, the pseudoscalar potential $A_0$) 
$
\vec A =(A_1(x,t),A_2(x,t),A_2(x,t)): {\mathbb R}^4\longrightarrow {\mathbb R}^3$, $\quad   A_0(x,t) : {\mathbb R}^4 \longrightarrow {\mathbb R}, \quad V_0  = V_0 ^* : {\mathbb R}^4 \longrightarrow M_4({\mathbb C})$.
We use the notation $V  ^* $ for the complex conjugate transpose of a matrix: $V ^* (x,t)=\ol{V (x,t)^T} $.  Hence, the Dirac equation can be written as follows, 
\[
  \dsp 
\left( {\gamma }^0  \left(   i  \partial_0 +A_0(x,t)\right)  +\sum_{\ell=1,2,3}{\gamma }^\ell \left( i e^{-Ht} \partial_\ell+A_\ell(x,t)\right)  +i \frac{3}{2}    H {\gamma }^0     -m{\mathbb I}_4  +{\gamma }^0 V_0 (x,t)\right)\psi=0  \,,
\]
with the notation
\[
{\alpha }^k={\gamma }^0{\gamma }^k, \quad ({\alpha }^k)^*={\alpha }^k,\quad k=1,2,3, 
\]
\[
\alpha ^1=\left(
\begin{array}{cccc}
 {\mathbb O}_2 & \sigma _1   \\
 \sigma _1 & {\mathbb O}_2  \\ 
\end{array}
\right),\quad \alpha ^2=\left(
\begin{array}{cccc}
{\mathbb O}_2 & \sigma _2   \\
 \sigma _2 & {\mathbb O}_2  \\ 
\end{array}
\right), \quad \alpha ^3=\left(
\begin{array}{cccc}
{\mathbb O}_2 & \sigma _3   \\
 \sigma _3 & {\mathbb O}_2  \\ 
\end{array}
\right),
\]
it can be also written as follows
\[
  \dsp 
\left(      \partial_0   + e^{-Ht}\sum_{\ell=1,2,3}\alpha ^\ell  \partial_\ell +\frac{3}{2}    H  {\mathbb I}_4    +im{\gamma }^0  -iV (x,t) \right)\psi=0 \,.
\]
In this paper, we study the semilinear Dirac equation, which describes the self-interacting field   
due to 
the nonlinear term $F=F(\psi )$ satisfying the following condition.
\medskip

\noindent
{\bf Condition (${\mathcal L}$)} The function $F=F(\psi ) \in C^3({\mathbb C}^4;{\mathbb C}^4)$   is Lipschitz continuous with exponent $\alpha  $ in the space $H_{(s)}({\mathbb R}^3) $,
that is, there exists  a constant $C>0$ such that 
\[
\| F(\psi _1)-F(\psi _2)\|_{H_{(s)}({\mathbb R}^3)} \leq C \|  \psi _1 - \psi _2 \|_{H_{(s)}({\mathbb R}^3)}
\left(\|  \psi _1   \|_{H_{(s)}({\mathbb R}^3)}^\alpha   +\|    \psi _2 \|_{H_{(s)}({\mathbb R}^3)}^\alpha  \right)\,. 
\]
The polynomial in $\psi$  vector-valued functions and the functions $F(\psi ) =(\gamma ^0 \psi, \psi) \gamma ^0 \psi $, $F(\psi ) =|\gamma ^0\gamma ^5 \psi|^\alpha   \psi $, $F(\psi ) =|\gamma ^0\gamma ^5 \psi|^\alpha  \gamma ^0 \psi $, 
$F(\psi ) = \pm|\psi |^{1+\alpha }$, $F(\psi ) = \pm|\psi |^{ \alpha } \psi $ 
are important examples of the Lipschitz continuous with exponent $\alpha  > 0$ in
the Sobolev space $H_{(s)}({\mathbb R}^3)$, $s \geq 3$, functions.

In the next theorem the mass $m$ is allowed to assume a complex value $m \in \C$, taking into account that  in the important cases of  $m=0 ,\pm iH$ the Dirac equation obeys  Huygens' principle \cite{Wunsch,JPh_2021}. 
Define the space
\[
X(R, s, \gamma ) :=\Big\{\varphi (x,t)\in C([0,\infty);H_{(s)}({\mathbb R}^3) )\,\Big|\, \| \varphi \|_X := \sup_{[0,\infty)  } e^{ \gamma   t} \| \varphi (x,t)\|_{H_{(s)}({\mathbb R}^3) } \leq R   \Big\}\,.
\]
For the potential $V$ we  write $V \in  {\mathscr{B}}_{(k;\ell)}$ that implies that all entries of the matrix $V$  belong to   the  space
\begin{eqnarray*} 
{\mathscr{B}}^{(\ell,k)}:= \Big\{v \in C_{t,x}^{\ell,k}  ([0,\infty)\times {\mathbb R}^3)  \,\Big|\, 
 \partial_t^j \partial_x^\alpha  v (x,t) \in L^\infty([0,\infty)\times {\mathbb R}^3)  , \,\forall \alpha ,\,|\alpha |\leq k,\,\,  
\forall j \leq \ell  \Big\}\,.
\end{eqnarray*}
\begin{theorem}
\label{T1.1}
Let $F=F(\psi ) \in C^3({\mathbb C}^4;{\mathbb C}^4)$  be the Lipschitz continuous with exponent $\alpha >0 $ in the space $H_{(s)}({\mathbb R}^3) $, $s\geq  3 $, function, and  the potential $V \in {\mathscr{B}}^{(s,0)} $ is   self-adjoint,
$V (x,t)= V ^*  (x,t) $. Assume  also that $2|\Im (m)|<3H$.

   Then there is 
$\varepsilon_0 >0 $ such that for all  
  $\varepsilon $ and $\psi_0 \in H_{(s)} $, $  \|\psi_0 \|_{(s)}\leq \varepsilon <\varepsilon_0 $,      the problem
\begin{equation}
\label{CPsmall}
\cases{  \dsp 
\left(      \partial_0   + e^{-Ht}\sum_{\ell=1,2,3}\alpha ^\ell  \partial_\ell +\frac{3}{2}    H  {\mathbb I}_4    +im{\gamma }^0 
-iV (x,t) \right)\psi=F(\psi ) \,,\cr 
\psi (x,0)=\psi_0 (x ) }
\end{equation}
  has a global solution $\psi \in X(2\varepsilon , s,   \frac{3}{2}     H  - |\Im (m)| ) $. The solution scatters to a free solution  as $t \to +\infty$.
\end{theorem}
We do not know if the condition $ 3    H  > 2|\Im (m)|  $ is necessary for the global existence of the solution to the equation of (\ref{CPsmall}).
Since the Dirac equation (\ref{LDE}) is non-invariant with respect to time inversion and its
solutions have different properties in different directions of time, we claim  in Theorem~\ref{T1.4} (Subsection~\ref{SS1}) only the asymptotic  behavior  of   the  solution at large positive time. In fact, the solution is asymptotically free under the following assumption:
 \[
4|\Im (m)|   +2|\Im (m)| \alpha   <3H\alpha\,.
\] 
In the case of $V (x,t)\equiv0$, the asymptotics at $t \to +\infty$ is written  via the explicit representation formulas, which are  derived in \cite{AP2020} for the solution of the Dirac equation in the de~Sitter spacetime.
 
\medskip

Concerning  small  amplitude global solutions of the nonlinear Dirac equation in the Minkowski space ($H=0$), one can consult    \cite{Shuji}, where one can also find 
previous references on that topic.   
Machihara,  Nakamura, Nakanishi, and Ozawa  \cite{Shuji} considered the nonlinear Dirac equation with the real mass and cubic nonlinearity $F(\psi )=-i\lambda (\gamma ^0 \psi , \psi )\gamma ^0 \psi $.   The  global well-posedness for small 
$H_{(1)}(\R^3)$  data with a slight regularity for angular variables  is proven in \cite{Shuji} by using the endpoint Strichartz estimates.
\medskip

Next we consider the Cauchy  problem with large data 
\begin{equation}
\label{NDE_CP}  
\cases{ \dsp\left(  i {\gamma }^0    \partial_0   +i e^{-Ht}\sum_{\ell=1,2,3}{\gamma }^\ell  \partial_\ell +i \frac{3}{2}    H {\gamma }^0     -m{\mathbb I}_4 +\gamma ^0 V(x,t)\right)\psi (x,t)  \cr 
\hspace{6cm}=F\left(\psi  ^*\gamma ^0 \psi\,,\,\psi  ^*\gamma ^0\gamma ^5 \psi \right)\psi (x,t), \quad t>0  ,\cr
\psi (x,0)=\Psi _0 (x)+\varepsilon \chi _0 (x).
}
\end{equation}
Here the function  $F=F(\xi ,\eta ) $, $F \in C^\infty({\mathbb R}^2;{\mathbb C}^4)$, has the form   
\begin{equation}
\label{F}
F\left( \psi  ^*\gamma ^0 \psi\,,\,\psi  ^*\gamma ^0\gamma ^5 \psi \right)= \alpha \left(\psi  ^*\gamma ^0 \psi\,,\,\psi  ^*\gamma ^0\gamma ^5 \psi \right){\mathbb I}_4 +i\beta \left(\psi  ^*\gamma ^0 \psi\,,\,\psi  ^*\gamma ^0\gamma ^5 \psi \right) \gamma ^5,\,\, 
\end{equation}
where $\alpha $ and $\beta $ are real-valued functions and
\begin{equation}
\label{Falphabeta} 
\alpha (\xi ,\eta ) =O(|\xi| +|\eta| ), \quad \beta  (\xi ,\eta ) =O(|\xi| +|\eta| ), \quad |\xi| +|\eta| \to 0\,. 
\end{equation}

The equation is a symmetric hyperbolic system, and the local existence of the solution is known (see, e.g., \cite{Taylor_III}). The local Cauchy problem for (\ref{NDE_CP}) is well posed  in $C^0([0,T];$ $(H_{(s)}({\mathbb R}^3))^4)  $, $s\geq 3$,  for some $T>0$ (see, e.g., \cite{Kato}). 

The main result of this paper is the following statement.  
\begin{theorem}
\label{T4.1}
Assume that $ m \in \R$, $H>0$,   
and the potential    $V \in  {\mathscr{B}}^{(\infty,\infty)} $ is  self-adjoint,
  $V^*(x,t)=V(x,t)$. Moreover, assume that 
\begin{equation}
\label{V}
V^T(x, t)\gamma ^2+\gamma ^2V (x, t)=0\,,
\end{equation}
while  $F$ takes the form  (\ref{F}) with (\ref{Falphabeta}). Assume also that the function $\Psi_0 =\Psi_0 (x)\in C_0^\infty ({\mathbb R}^3;{\mathbb C}^4)$ satisfies the Lochak-Majorana condition
\begin{equation}
\label{LMC}
\rho ^2(\Psi_0(x)):= | \Psi_0 ^*(x) \gamma ^0\Psi_0 (x)|^2  + |\Psi_0 ^*(x) \gamma ^0\gamma ^5\Psi_0(x)|^2  = 0
\quad \mbox{\it for all }\quad x \in \R^3\,.  
\end{equation}

Then for $\chi _0 \in C_0^\infty ({\mathbb R}^3;{\mathbb C}^4) $  
there exists an  $\varepsilon _0>0 $ such that the Cauchy problem (\ref{NDE_CP}) with  $ 0<\varepsilon <\varepsilon _0$ has a unique solution $\psi =\psi (x,t) $ such that as a function of time  $\psi (t) \in  C_0 ^\infty ({\mathbb R}^3;{\mathbb C}^4)  $ for all $t \in (0,\infty)$. The solution scatters to the  solution of the free Dirac equation as $t \to +\infty$.
\end{theorem}
Thus, the global in time solutions exist  for the initial data from the small neighborhood of the unbounded conic set of functions satisfying  the Lochak-Majorana condition. The last condition is independent of the mass $m$,  the value of the positive Hubble constant $H>0$, and, consequently,  the curvature of spacetime. Another   aim of the present article is  to gain  insight into the asymptotic behavior of an open set of large data solutions. In order to  write an asymptotic at large time for the solutions given by  Theorems~\ref{T1.1},\ref{T4.1}, we use in Theorems~\ref{T1.4},~\ref{T4.1A} the 
 explicit representation formulas for the solution of the Dirac equation in the de~Sitter spacetime, which are  obtained in \cite{AP2020}.

\begin{remark}
If the potential  $V$ is independent of $x$, then the  condition of the boundedness of the $V$ can be dropped.  
\end{remark}
The global existence of large amplitude solutions for the nonlinear massless Dirac equation  in the Minkowski space  was considered by Bachelot~\cite{Bachelot}. The only smallness assumption in \cite{Bachelot} was on the chiral invariant related to the Lochak-Majorana condition. In \cite{Bachelot} the asymptotic behavior of the solution, particularly the equipartition of energy and 
the decay of Lorentz-invariant products, were also studied.  Bachelot in  \cite{Bachelot} developed the approach  that appeals to the estimates obtained by the replacing the generators of the Poincar\'e  group with the Fermi operators. That method  requires certain regularity of the data in order to be applied.

D'Ancona and Okamoto \cite{Piero} studied a massless cubic Dirac equation in the Minkowski space 
\[
  \dsp 
\left(     i \partial_0   + i \sum_{\ell=1,2,3}\alpha ^\ell  \partial_\ell       +V (x) \right)\psi= \langle \gamma ^0 u,u \rangle \gamma ^0 u \,,
\]
perturbed by a large potential with almost critical regularity. They  proved global existence and scattering for small initial data in $H_{(1)}$ with additional angular regularity. The main tool was the endpoint Strichartz estimate for the perturbed Dirac flow that was proved by  the arguments of \cite{Federico-Piero}. In particular, the result of \cite{Piero} covers the case of spherically symmetric data with small $H_{(1)}$ norm. In the absence of the magnetic potential $\vec A$, and when   $0$ is not a resonance for $ i\sum_{\ell=1,2,3}\alpha ^\ell  \partial_\ell     +V (x)$, and the potential $V=A_0 \gamma^0+V_0$ satisfies (\ref{V}), D'Ancona and Okamoto  proved the global existence and scattering for large initial data having a small chiral component, related to the Lochak-Majorana condition.

In \cite{Candy}, Candy and Herr proved that the Cauchy problem for the cubic Dirac equation in the Minkowski space is globally well-posed and that its solutions scatter to free solutions as $t \to \pm \infty$.

The rest of this paper is organized as follows.  In Section~\ref{S1}, we prove Theorem~\ref{T1.1} by derivation of the energy estimate (subsection~\ref{SSEE})  and fixed point argument (subsection~\ref{SFPA}).
 The asymptotics on the positive half-line of time and the representation of the solution of the free  Dirac equation in de~Sitter space are given in Theorem~\ref{T1.4} (subsection~\ref{SS1}). In Section~\ref{S2}, we analyze  the Lochak-Majorana condition in the de~Sitter space and obtain its time-evolution. In Section~\ref{S3}, we prove   Theorem~\ref{T4.1} except the asymptotics part that follows from Theorem~\ref{T4.1A} of  
 Section~\ref{S4}. There, we also give the asymptotics at infinity for the solutions of Theorem~\ref{T4.1}. 
In  Section~\ref{S5}, we show the blow-up result for the large data solution of the Dirac equation in expanding (Theorem~\ref{TBUH+},    $H>0$) and contracting (Theorem~\ref{T5.3}, $H<0$) universes. Finally, in  Section~\ref{SS5.4}, we prove that the classical solutions of the semilinear Dirac equation obey the finite propagation speed property.

\section{Small data global existence}
\label{S1}
\setcounter{equation}{0}

\subsection{Proof of Theorem~\ref{T1.1}. Energy estimate}
\label{SSEE}

First we develop the energy estimates for the solution of the equation. 
Denote the Dirac operator in the de~Sitter space by
\[
{\mathscr{D}}_{dS}(x,t, \partial):=       \partial_0   + e^{-Ht}\sum_{\ell=1,2,3}\alpha ^\ell  \partial_\ell +\frac{3}{2}    H  {\mathbb I}_4    +im{\gamma }^0 -iV (x,t)
\] 
and
\begin{equation}
\label{dplusdminus}
\delta _+:= - 3    H  + 2|\Im (m)|,\qquad \delta _-:= - 3    H  - 2|\Im (m)|\,.
\end{equation}

\begin{lemma}
\label{L1.2}
Assume that $H \in \R$,    $f \in C([0,\infty);(H_{(k)} ({\mathbb R}^3))^4 )$, and that the potential $V\in {\mathscr{B}}_{(0,k)}$  is self-adjoint,  $V (x,t)= V (x,t)^* $. 
Then for the solution $\psi =\psi (x,t) $ of the equation
\begin{equation}
\label{LinEq}
{\mathscr{D}}_{dS}(x,t, \partial)\psi(x,t)=f(x,t)   
\end{equation}
in the Sobolev space $H_{(k)}({\mathbb R}^3)$ one has
\begin{eqnarray*}
\|\psi (t) \|_k 
& \leq  & 
C_k  e^{  \frac{1}{2}\delta_+ (t-s)} \|\psi (s) \|_k +
 C_k    e^{  \frac{1}{2}\delta_+ t} \int_s^t  e^{-\frac{1}{2}\delta _+\tau }\| f (\tau )\|_k \,d\tau  \quad \mbox{for all}\quad s\leq t\,,\\
\|\psi (t ) \|_k
& \leq  &
C_ke^{ \frac{1}{2}\delta_-(t-s)}\|\psi (s ) \|_k  +  C_k e^{ \frac{1}{2}\delta_- t}\int_t^s e^{-\frac{1}{2}\delta _-\tau }\| f (\tau )\|_k \, d\tau \quad \mbox{for all}\quad t\leq s \,.
\end{eqnarray*}
\end{lemma}
\medskip

\ndt
{\bf Proof.} 
Consider the energy integral
\begin{eqnarray*}
E(t)
& = &
\int_{{\mathbb R}^3}|\psi (x,t)|^2\,dx \,.
\end{eqnarray*} 
Then, by the finite speed propagation property (see Section~\ref{SS5.4})
\[
\frac{d}{d t} E(t)
=
  \int_{{\mathbb R}^3}\left[  - 3    H   |\psi (x,t)|^2   + 2\Im (m) \psi^* (x,t){\gamma }^0   \psi (x,t)   \right] \,dx 
 +\int_{{\mathbb R}^3}2\Re \left[  f^* (x,t)  \psi (x,t) \right] \,dx\,.
\]
Here the following identities  have been used
\[
 (\partial_k  \psi^* (x,t)){\alpha }^k   \psi (x,t) +\psi^* (x,t) {\alpha }^k  \partial_k  \psi (x,t)  
 = 
\partial_k(  \psi^* (x,t) \alpha^k\psi (x,t))\,,
\]    
\[
\int_{{\mathbb R}^3} \left[  (\partial_k  \psi^* (x,t)){\alpha }^k   \psi (x,t) +\psi^* (x,t) {\alpha }^k  \partial_k  \psi (x,t)   \right]  \,dx 
 = 
0\,,
\]
since $({\alpha }^k)^*={\alpha }^k$  ($k=1,2,3 $). The assumption  $H\in \R$ implies
\begin{eqnarray*}
 \left(   - 3    H  - 2|\Im (m)|\right)   |\psi (x,t)|^2 
& \leq &
 - 3    H   |\psi (x,t)|^2   + 2\Im (m) \psi^* (x,t){\gamma }^0   \psi (x,t)  \\
& \leq  &
 \left(   - 3    H  + 2|\Im (m)|\right)   |\psi (x,t)|^2 \,.
\end{eqnarray*}
Then
\[
\delta _-   |\psi (x,t)|^2 \leq 
   - 3    H   |\psi (x,t)|^2   + 2\Im (m) \psi^* (x,t){\gamma }^0   \psi (x,t)  \leq 
\delta _+  |\psi (x,t)|^2 
\]
and
\begin{equation}
\label{DIN}
\delta _-  E(t)
 +\int_{{\mathbb R}^3}2\Re \left[ f^* (x,t) \psi (x,t) \right] \,dx
  \leq  
\frac{d}{d t} E(t)\leq 
\delta _+  E(t)
 +\int_{{\mathbb R}^3}2\Re \left[ f^* (x,t) \psi (x,t) \right] \,dx   \,.
\end{equation}
In particular, the right-hand side gives
\[
\frac{d}{d t} E(t)
 \leq  
\delta _+  E(t)
 +\int_{{\mathbb R}^3}2\Re \left[ f^* (x,t) \psi (x,t) \right]  \,dx
\]
that leads to
\[
\frac{d}{d t} \left( E(t)e^{-\delta _+t} \right)
 \leq 
 2e^{-\delta _+t}\| f (x,t)\| \| \psi (x,t) \|\,.
\]
We integrate it from $s$ to $t$:
\[
 E(t)e^{-\delta _+t} 
  \leq   
E(s)e^{-\delta _+s}+
 2\int_s^te^{-\delta _+\tau }\| f (\tau )\| \| \psi (\tau ) \|\, d\tau, \quad s\leq t\,.
\]
If we fix $s$ and denote
\[
y(t) :=\max_{\tau \in [s,t]} e^{- \frac{1}{2}\delta_+ \tau}\|\psi (\tau ) \|_k, \quad  y^2(t) =\max_{\tau \in [s,t]} e^{ -\delta_+ \tau}\|\psi (\tau ) \|_k^2\,,  
\]
then for $k=0$ the inequality 
\[
y^2(t)
  \leq   
y^2(s)+
 2y(t)\int_s^t  e^{-\frac{1}{2}\delta _+\tau }\| f (\tau )\|_k \,d\tau 
\]
yields 
\[
y^2(t)
  \leq  
c^2 y^2(s)+
 c^2  \left(  \int_s^t  e^{-\frac{1}{2}\delta _+\tau }\| f (\tau )\|_k \,d\tau \right)^2\,.
\]
Consequently, for $k=0$ we have obtained
\[
e^{- \frac{1}{2}\delta_+ t}\|\psi (t) \|_k 
  \leq    
c  e^{- \frac{1}{2}\delta_+ s}\|\psi (s) \|_k +
 c     \int_s^t  e^{-\frac{1}{2}\delta _+\tau }\| f (\tau )\| _k\,d\tau \,,\quad s\leq t\,,
\] 
that is, the first inequality of the lemma.

Next, we choose the left-hand side of (\ref{DIN})
\[
\delta _-  E(t)
 +\int_{{\mathbb R}^3}2\Re \left[ f^* (x,t) \psi (x,t) \right] \,dx
  \leq   
\frac{d}{d t} E(t)  
\] 
and rewrite it as follows:
\[
e^{-\delta _-t}\int_{{\mathbb R}^3}2\Re \left[ f^* (x,t) \psi (x,t) \right] \,dx
  \leq  
\frac{d}{d t} \left( E(t) e^{-\delta _-t} \right)  \,.
\]
We integrate the last inequality from $t$ to $s$, $t<s$: 
\[
\int_t^s  e^{-\delta _-t}\int_{{\mathbb R}^3}2\Re \left[ f^* (x,t) \psi (x,t) \right] \,dx
  \leq  
E(s) e^{-\delta _-s} -  E(t) e^{-\delta _-t}  \,.
\]
It follows
\[
- 2\int_t^s e^{-\delta _+\tau }\| f (\tau )\| \| \psi (\tau ) \|\, d\tau 
 \leq 
E(s) e^{-\delta _-s} -  E(t) e^{-\delta _-t}  
\]
and
\[
 E(t) e^{-\delta _-t}   
 \leq 
E(s) e^{-\delta _-s} + 2\int_t^s e^{-\delta _-\tau }\| f (\tau )\| \| \psi (\tau ) \|\, d\tau \,,\quad t\leq s \,.
\]
If we fix $s$ and denote
\[
y_-(t) :=\max_{\tau \in [t,s]} e^{- \frac{1}{2}\delta_- \tau}\|\psi (\tau ) \|_k, \quad  y^2_-(t) =\max_{\tau \in [t,s]} e^{ -\delta_- \tau}\|\psi (\tau ) \|_k^2\,,  
\]
then for $k=0$ the estimate
\[
 y^2_-(t) 
 \leq 
y^2_-(s) + 2 y_-(t) \int_t^s e^{-\frac{1}{2}\delta _-\tau }\| f (\tau )\| \, d\tau \,,\quad t\leq s\,, 
\]
implies
\[
 y _-(t) 
\leq 
cy_-(s)  +  c \int_t^s e^{-\frac{1}{2}\delta _-\tau }\| f (\tau )\| \, d\tau \,,\quad t\leq s\,, 
\]
that yields   for $k=0$ 
\[
e^{- \frac{1}{2}\delta_- t}\|\psi (t ) \|_k
\leq 
ce^{- \frac{1}{2}\delta_-s}\|\psi (s ) \|_k  +  c \int_t^s e^{-\frac{1}{2}\delta _-\tau }\| f (\tau )\|_k \, d\tau \,,\quad t\leq s \,,
\]
that is,  the second inequality of the lemma.  For every $k>0$ the inequalities follow  from the case of $k=0$ by differentiation. The lemma is proved. \qed

\medskip

\subsection{Proof of Theorem~\ref{T1.1}}
\label{SFPA}

\ndt
We are going to apply the Banach fixed-point theorem. 
Denote by $S(t,s) $ the propagator (fundamental solution for the Cauchy problem),  that is, an operator-valued  solution of 
the problem 
\begin{eqnarray}
\label{Prop}
\hspace{-0.9cm} &  &
\cases{  \dsp 
{\mathscr{D}}_{dS}(x,t, \partial)S(t,s)=0\,,\quad t , s \in \R\,,\cr 
S(s,s)=I \,(\mbox{identity operator})\,.
}
\end{eqnarray}
Then the solution of the problem  
\begin{eqnarray*}
&  &
\cases{  \dsp  
{\mathscr{D}}_{dS}(x,t, \partial)\psi (x,t)=f(x,t)\,,\quad t \geq s\,,\cr
\psi (x,s)=\psi_0 (x )\,,
} 
\end{eqnarray*}
is given by Duhamel's principle
\[
\psi (x,t)= S(t,s) \psi_0 (x )+\int_s^t S(t,\tau )f(x,\tau)\,d \tau\,.
\]
It is known (see, e.g., \cite{Kato,Taylor_II}) that for $ \psi_0 \in H_{(k)}({\mathbb R}^3)$ and $f \in C([0,\infty)  ; H_{(k)}({\mathbb R}^3))$, $k>5/2$, the unique solution $\psi  \in C([0,\infty)  ; H_{(k)}({\mathbb R}^3))\cap C^1([0,\infty)  ; H_{(k-1)}({\mathbb R}^3))$ exists.  

Next, we define the operator $\mathcal S$ by
\[
{\mathcal  S} \psi(x,t) :=S(t,0) \psi _0 ( t)+\int_0^t S(t,\tau )F(\psi(x,\tau))\,d \tau\,.
\]
We  are going to prove that ${\mathcal  S}$ is a contraction such that   
\begin{eqnarray*}
&  &
{\mathcal  S} : X(R, s, \gamma ) \longrightarrow  X(R, s, \gamma )
\end{eqnarray*}
for sufficiently small $ \varepsilon _0$ and $R$.

According to Lemma~\ref{L1.2}, with $k>3/2$ we have 
\[
\|\psi (t) \|_k 
\leq 
c  e^{  \frac{1}{2}\delta_+ (t-s)} \|\psi (s) \|_k +
 c    e^{  \frac{1}{2}\delta_+ t} \int_s^t  e^{-\frac{1}{2}\delta _+\tau }\| f (\tau )\|_k \,d\tau  \,,\quad s<t\,.
\]
Set $s=0$ and $\delta := - \frac{1}{2}\delta_+=\frac{1}{2}( 3    H -  2|\Im (m)|)>0$ in the last inequality, then by using the condition  (${\mathcal L}$) we obtain 
\begin{eqnarray*}
\|\psi (t) \|_k
& \leq  &
c e^{-\delta  t}\|\psi (0) \|_k+  c e^{-\delta  t}\int_0^t  e^{\delta  \tau }   \| F(\psi (\tau )) \|_k       \,d\tau \\
& \leq  &
c e^{-\delta  t}\|\psi (0) \|_k+ c e^{-\delta  t}\int_0^t  e^{\delta  \tau }    \| \psi(\tau ) \|_{k}^{1+ \alpha }  \,d\tau,\quad t >0 \,.
\end{eqnarray*}
Hence, 
\begin{eqnarray*}
\sup_{t \in [0,\infty) } e^{ \delta  t} \| \psi (t) \|_{k} 
& \leq  &
 c\|\psi (0) \|_{(k)}+  c\left( \sup_{t \in [0,\infty) }e^{ \delta  t} \| \psi (t) \|_{k} \right)^{1+\alpha }\int_0^t e^{- \alpha \delta  \tau }   \,d\tau \\ 
& \leq  &
c\|\psi (0) \|_{(k)}+  c(\alpha \delta )^{-1} (1- e^{-\alpha \delta t}) \left( \sup_{t \in [0,\infty) } e^{ \delta  t} \| \psi (t) \|_{k} \right)^{1+\alpha } \\ 
& \leq  &
C\|\psi (0) \|_{(k)}+  C(\alpha \delta )^{-1} \left( \sup_{t \in [0,\infty) } e^{ \delta  t} \| \psi (t) \|_{k} \right)^{1+\alpha }\,.
\end{eqnarray*}
Thus, the last inequality proves that the operator ${\mathcal  S}$  maps 
\[
X(R, k, \delta  ) :=\Big\{\varphi (x,t)\in C([0,\infty);H_{(k)})\,\Big|\, \| \varphi \|_X := \sup_{[0,\infty)  } e^{ \delta  t} \| \varphi (x,t)\|_{H_{(k)}} \leq R   \Big\}
\]
 into itself provided that  the initial function $ \psi _0 (x) \in H_{(k)}$, $\|\psi _0\|_{H_{(k)}}< \varepsilon $, and  that the numbers $\varepsilon  $ and $R$ are sufficiently small, namely, if 
\[
C \varepsilon +  C\frac{1}{\alpha \delta }R^{1+ \alpha }  < R \,.
\]
In order to verify a contraction property, we write
\[
{\mathcal  S} \psi_1(x,t) - {\mathcal  S} \psi_2(x,t) = \int_0^t S(t,\tau )\left(  F(\psi_1(x,\tau)) -F(\psi_2(x,\tau)) \right) \,d \tau
\] 
and use Lemma~\ref{L1.2} and the condition ($\mathcal L$) to estimate the norm 
\begin{eqnarray*} 
\|{\mathcal  S} \psi_1(x,t) - {\mathcal  S} \psi_2(x,t)\|_k  
& \leq  & 
c    e^{  -\delta t} \int_0^t  e^{\delta \tau }\|   F(\psi_1( \tau)) -F(\psi_2( \tau)) \|_k  \,d \tau \\ 
& \leq  & 
 C    e^{   -\delta  t} \int_0^t  e^{\delta \tau }
\|  \psi _1 (\tau)- \psi _2(\tau) \|_k
\left(\|  \psi _1 (\tau)  \|_k^\alpha   
+\|    \psi _2 (\tau)\|_k^\alpha  \right)\,d \tau \,.
\end{eqnarray*} 
It follows
\begin{eqnarray*} 
&   & 
\sup_{t \in [0,\infty) } e^{ \delta  t}\|{\mathcal  S} \psi_1(x,t) -{\mathcal  S} \psi_2(x,t)\|_k \\ 
& \leq  &
 C   \left( \sup_{\tau  \in [0,\infty) }  e^{\delta \tau }
\|  \psi _1 (\tau)- \psi _2(\tau) \|_k \right)  \int_0^t 
e^{-\alpha \delta \tau }e^{\alpha \delta \tau }\left(\|  \psi _1 (\tau)  \|_k^\alpha   
+\|    \psi _2 (\tau)\|_k^\alpha  \right)\,d \tau \\ 
& \leq  &
 C   \left( \sup_{\tau  \in [0,\infty) }  e^{\delta \tau }
\|  \psi _1 (\tau)- \psi _2(\tau) \|_k \right)  \left( \sup_{\tau  \in [0,\infty) }  e^{\delta \tau }
\left(\|  \psi _1 (\tau)  \|_k   
+\|    \psi _2 (\tau)\|_k  \right)\right)^\alpha   \int_0^t 
e^{-\alpha \delta \tau }\,d \tau \\ 
& \leq  &
 C_1   \left( \sup_{\tau  \in [0,\infty) }  e^{\delta \tau }
\|  \psi _1 (\tau)- \psi _2(\tau) \|_k \right)  \left( \sup_{\tau  \in [0,\infty) }  e^{\delta \tau }
\left(\|  \psi _1 (\tau)  \|_k   
+\|    \psi _2 (\tau)\|_k  \right)\right)^\alpha  \\ 
& \leq  &
 C_1   d(\psi _1 ,\psi _2 )  
2^\alpha R^\alpha \,.
\end{eqnarray*} 
Then we choose  $R$ such that $C_12^\alpha R^\alpha <1 $. Banach fixed-point theorem completes the proof of Theorem~\ref{T1.1}. 
 \qed

\subsection{Large  time asymptotics }  
\label{SS1}

For the given  initial function  $\psi _0(x) $ we want to show that there are the solution $\psi (x,t) $ of
the problem
\begin{equation}
\label{1.7DE}
\cases{ \dsp 
{\mathscr{D}}_{dS}(x,t, \partial)\psi (x,t) =F(\psi  (x,t) ) \,,\cr  \psi (x,0)=\psi_0 (x )\,,
}
\end{equation}
and the solution $\psi ^{+}(x,t) $ of the free Dirac equation
\begin{eqnarray}
\label{DFree}
\cases{ \dsp 
{\mathscr{D}}_{dS}(x,t, \partial)\psi ^{+}(x,t) =0\,,\cr 
\psi^{+} (x,0)=\psi^{+}_0 (x )\,,
}
\end{eqnarray} 
such that 
\begin{equation}
\label{Splus}
\lim_{t \to +\infty}    \left\|  \psi (x,t)-    \psi^{+} (x,t) \right\|_{(H_{(k)}({\mathbb R}^3))^4}=0\,.
\end{equation}
From the viewpoint of
scattering theory the function $\psi (x,t)$ has  free  asymptotics as $t \to +\infty$. 
The mapping ${\bf \mathscr{S}_+}\,:\,\psi_0  \longmapsto  \psi^{+}_0 $ is  related  to the {\it   wave  operator $W_+$}  of the scattering theory   if (\ref{Splus}) is fulfilled (see, e.g., \cite{Strauss,Yafaev}).

We are going to prove the existence of the  operator ${\bf \mathscr{S}_+} $. 
Let  $S(t,s) $ be the propagator,  that is, the  solution of 
the problem (\ref{Prop}), then
\[
\psi (x,t)=S(t,s)\psi (x,s)\,,\qquad t,s \in \R,\quad x\in \R^3\,.
\]
The  operator $S(t,0) $  is written in the explicit form in \cite{AP2020}.
To find for  the operator $S(t,s) $  a similar  explicit form one can use the symmetry property  of the Dirac operator and of the de~Sitter spacetime (see, \cite[Sec.142]{Tolman}) . 

With the aid of the operator  $S(t,s) $, we look for the solution of (\ref{1.7DE}) via the following integral equation
\[
 \psi (x,t)=   \psi^{+} (x,t) - \int_{t }^\infty S(t,\tau)F(\psi  (x,\tau))\, d \tau  \,, 
\]
provided that the  integral is convergent. For the initial condition we have  
\begin{equation}
\label{1.7}
\psi^{+}_0 (x ) = \psi_0 (x )   + \int_{0}^\infty S(0,\tau)F(\psi  (x,\tau))\, d \tau \,.
\end{equation}
To prove the  existence of the operator  ${\bf \mathscr{S}_+} $ it suffices  a  convergence of the  integral of (\ref{1.7}).

\begin{lemma}
\label{L2.3}
Let $F=F(\psi ) \in C^3({\mathbb C}^4;{\mathbb C}^4)$  be the Lipschitz continuous with exponent $\alpha >0 $ in the space $H_{(k)}({\mathbb R}^3) $, $k\geq  3 $, function. 
Assume that the potential  $V\in {\mathscr{B}}_{(0,k)}$ is   self-adjoint,  $V (x,t)= V (x,t)^* $, 
$4|\Im (m)|   +2|\Im (m)| \alpha   <3H\alpha  $,  
and the function $ \psi   \in C([0,\infty);(H_{(k)}({\mathbb R}^3))^4) $   solves the problem (\ref{CPsmall}). 

Then the  limit
\begin{eqnarray}
\label{2.12}
\lim_{t \to +\infty}\int_{0}^t S(0,\tau)F(\psi  (x,\tau))\, d \tau    
\end{eqnarray}
exists in the space  $(H_{(k)}({\mathbb R}^3))^4 $.
\end{lemma}
\medskip

\noindent
{\bf Proof.} According to Lemma~\ref{L1.2}  we have 
\begin{eqnarray*}  
\|S(0,\tau )F(\psi  (x,\tau)) \|_k
& \leq  &
ce^{ \frac{1}{2}\delta_-( -\tau )}\|F(\psi  (x,\tau))\|_k \\ 
& \leq  &
ce^{ \frac{1}{2}\delta_-( -\tau )}\|\psi  (x,\tau)\|_k ^{1+\alpha}  \,,\quad 0<\tau  .
\end{eqnarray*} 
Then the existence of the limit follows from 
\begin{eqnarray*}
\|S(0,\tau )F(\psi  (x,\tau)) \|_k 
& \leq  &
ce^{ \frac{1}{2}\delta_-( -\tau )}\|\psi  (x,\tau)\|_k ^{1+\alpha}  \\
& \leq  &
c_1e^{- \frac{1}{2}(- 3    H  - 2|\Im (m)|) \tau +(- \frac{1}{2}(3    H - 2|\Im (m)|)\tau )(1+\alpha)} \| \psi (x,0 ) \|_k^ {1+\alpha}  \,,
\end{eqnarray*}
with $\delta _+ $ and $\delta _- $ of (\ref{dplusdminus}). 
According to the condition of the lemma 
\[
- \frac{1}{2}(- 3    H  - 2|\Im (m)|)  +(- \frac{1}{2}(3    H - 2|\Im (m)|) )(1+\alpha)
  <0
\]
and, consequently, the limit (\ref{2.12}) exists. The lemma is proved.
\qed 
\smallskip

Next we give an explicit representation formula for the solution  
$\psi ^{+}(x,t)$ that will be used in the theorem on the existence of the    operator   ${\bf \mathscr{S}_+} $ .
We define forward and backward light cones   as the boundaries of  $D_+(x_0, t_0)$ and $D_-(x_0, t_0)$, respectively, where
\begin{equation}
\label{1.3D} 
D_{\pm}\left(x_{0}, t_{0}\right) :=\left\{(x, t) \in {\mathbb R}^{3+1}\, \left| \right.\,
\left|x-x_{0}\right| \leq \pm\left(\phi (t) -\phi (t_{0})  \right)\right\}\,,
\end{equation}
and  $\phi (t):= (1-e^{-Ht} )/H$ is a distance function.  
For $(x_0, t_0) \in {\mathbb R}^3\times {\mathbb R}$, $M \in {\mathbb C}$, $r = |x- x_0 | /H $,  we follow \cite{Yag_Galst_CMP,AP2020} and define the function 
\begin{eqnarray*}
E(r,t;0,t_0;M)
& :=  &
4^{-\frac{M}{H}} e^{ M  (  t_0+ t)} \left(\left(e^{-H t_0}+e^{-H t}\right)^2-(H r)^2\right)^{\frac{M}{H}-\frac{1}{2}}  \nonumber \\
&  &
\times  F \left(\frac{1}{2}-\frac{M}{H},\frac{1}{2}-\frac{M}{H};1;\frac{\left(e^{-H t}- e^{-H t_0}\right)^2-(r H)^2}{\left(e^{-H t}+e^{-Ht_0}\right)^2-(r H)^2}\right)  \,,
\end{eqnarray*}
where   $(x,t) \in D_+ (x_0,t_0)\cup D_- (x_0,t_0) $  and $F\big(a, b;c; \zeta \big) $ is the hypergeometric function (see, e.g., \cite{B-E}). 
Denote  
\[
M_+=   \frac{1}{2}H  + im    ,\quad M_-=   \frac{1}{2}H  - im \,.
\]
Let $e^{H\cdot } $  be  the operator of multiplication by the function $e^{Ht } $. 
Theorem~0.2~\cite{AP2020}  gives the representation formula  for the solutions of the Cauchy problem. In order to formulate it, we need the operator ${\cal G}(x,t,D_x;M ) $ defined by 
\begin{eqnarray*}
&  &
{\cal G}(x,t,D_x;M )[f](x,t)  \\
& = &
2   \int_{ 0}^{t} db
  \int_{ 0}^{\phi (t)- \phi (b)}  E(r,t;0,b;M) \int_{{\mathbb R}^n} {\mathcal E}^w (x-y,r) f  (y,b) \,d y  \, dr  
,   \quad f \in C_0^\infty({\mathbb R}^{4}),
\end{eqnarray*}
where ${\mathcal E}^w (x,r) $ is a fundamental solution of the Cauchy problem
\[
\cases{ 
 v_{tt}-  \Delta  v =0, \quad x \in {\mathbb R}^3  \,,\quad t \in {\mathbb R}\,, \cr
 v(x,0)= \varphi (x), \quad v_t(x,0)=0\,,\quad x \in {\mathbb R}^3 \,,  
}
\]
 in the Minkowski space, that is, for $n=3$ 
  (see, e.g., \cite{Shatah})  
\[ 
{\mathcal E}^{w}(x, t) :=\frac{1}{4\pi } \frac{\partial}{\partial t}
\frac{1}{t} \delta(|x|-t)\,.
\]
The distribution $\delta(|x|-t)$ is defined
 by
\[
\langle \delta(|\cdot|-t), \psi(\cdot)\rangle  =\int_{|x|=t} \psi(x)\, d x \quad \mbox{\rm for } \quad \psi \in C_{0}^{\infty}\left({\mathbb R}^{3}\right).
\] 
The Kirchhoff's formula gives
\[
 v(x,t)= \frac{\partial}{\partial t}V(x,t), \quad \mbox{\rm where}\quad V(x,t)=  
\frac{t}{4\pi }\int_{|y|=1}\varphi (x+ty)\,dS_y\,.
\]
We also need the kernel function
\[
K_1(r,t;M)
 := 
E(r,t;0,0;M)   
\]
and the operator ${\cal K}_1(x,t,D_x;M)$, which is  defined  as follows:
\begin{equation}
\label{K1OPER}
{\cal K}_1(x,t,D_x;M) [\varphi] (x,t) = 
 2\int_{0}^{\phi (t) } 
  K_1( s,t;M)  \int_{{\mathbb R}^n} {\mathcal E}^w (x-y,s)  \varphi  (y) \,d y\, ds\,, \quad \varphi \in C_0^\infty({\mathbb R}^3),
\end{equation}
or
\[
{\cal K}_1(x,t,D_x;M) [\varphi] (x,t)
 = 
 2\int_{0}^{\phi (t) } 
  K_1( s,t;M) \, ds  \frac{\partial}{\partial s}
\frac{s}{4\pi }\int_{|y|=1}\varphi (x+sy)\,dS_y\,, \quad \varphi \in C_0^\infty({\mathbb R}^3).
\]

According to Theorem~0.2~\cite{AP2020}, 
the  solution to the Cauchy problem
\[
\cases{\dsp \left(i {\gamma }^0  \partial_0+i  e^{-Ht}\sum_{k=1,2,3}{\gamma }^k \partial_k  + i       \frac{3}{2}  H  {\gamma }^0   -m{\mathbb I}_4\right)\Psi (x,t)=F(x,t)\,,\cr 
\Psi (x,0)= \Phi   (x ) \,,
}
\] 
where $m \in {\mathbb C}$, is given by the following formula
\begin{eqnarray*} 
\Psi (x,t) 
& = &
 - e^{-Ht}\left( i\gamma ^0 \partial_0+  ie^{-Ht} \sum_{k=1,2,3}\gamma ^k \partial_k- i\frac{H}{2}\gamma ^0+m{\mathbb I}_4\right) \\
&  &
\times \left [   \left (
   \begin{array}{cccc}
 {\cal G}(x,t,D_x;M_+){\mathbb I}_2& {\mathbb O}_2   \\ 
  {\mathbb O}_2 & {\cal G}(x,t,D_x;M_-){\mathbb I}_2   \\ 
   \end{array}
   \right )[e^{H\cdot }F ](x,t) \right.\\
&  &
\left. +i\gamma ^0 \left (
   \begin{array}{cccc}
 {\cal K}_1(x,t,D_x;M_+){\mathbb I}_2 & {\mathbb O}_2  \\ 
 {\mathbb O}_2&  {\cal K}_1(x,t,D_x;M_-){\mathbb I}_2   \\ 
   \end{array}
   \right )[\Phi   ](x,t)\right ]  \,.
 \end{eqnarray*}

The next theorem states the existence of the  operator ${\mathscr{S}}^+\,:\,\psi_0  \longmapsto  \psi^{+}_0 $.
\begin{theorem}
\label{T1.4}
Let $F=F(\psi ) \in C^3({\mathbb C}^4;{\mathbb C}^4)$  be the Lipschitz continuous with exponent $\alpha >0 $ in the space $H_{(k)}({\mathbb R}^3) $, $k\geq  3 $, function.
Assume that the potential  $V\in {\mathscr{B}}_{(0,k)}$ is   self-adjoint,  $V (x,t)= V (x,t)^* $,
\[
4|\Im (m)|   +2|\Im (m)| \alpha   <3H\alpha \,.
\]
For every solution $ \psi  \in C([0,\infty);(H_{(k)}({\mathbb R}^3))^4) $ of the semilinear Dirac equation in the de~Sitter spacetime (\ref{1.7DE}), let
$ \psi^{+}_0 (x ) $ be  the function   given by
\[
\psi^{+}_0 (x ) = {\bf \mathscr{S}^+}\psi_0 (x )= \psi_0 (x )   + \int_{0}^\infty S(0,\tau)F(\psi  (x,\tau))\, d \tau \,,
\]
where $\psi_0 (x ) :=\psi  (x,0) $. 
Then the solution $ \psi^+ (x,t)$ of the Cauchy problem for the  free Dirac equation (\ref{DFree}) 
 satisfies  (\ref{Splus}) 
and  ${\mathscr{S}}^+\,:\,\psi_0  \longmapsto  \psi^{+}_0 $ is the  continuous    operator  in $H_{(k)}({\mathbb R}^3)$. 

Moreover, if $V(x,t)=0$, then  
\begin{eqnarray}
\label{2.20} 
\psi^+ (x,t) 
& = &
 - e^{-Ht}\left( i\gamma ^0 \partial_0+  ie^{-Ht} \sum_{k=1,2,3}{\alpha }^k  \partial_k- i\frac{H}{2}\gamma ^0+m{\mathbb I}_4\right) \\
&  &
\times \left [   i\gamma ^0 \left (
   \begin{array}{cccc}
 {\cal K}_1(x,t,D_x;M_+){\mathbb I}_2 & {\mathbb O}_2  \\ 
 {\mathbb O}_2&  {\cal K}_1(x,t,D_x;M_-){\mathbb I}_2   \\ 
   \end{array}
   \right )[\psi^{+}_0 ]\right ]  \,.\nonumber
 \end{eqnarray} 
\end{theorem}
\medskip

\noindent
{\bf Proof.} According to Lemma~\ref{L2.3} the function $\psi^{+}_0 (x ) $ exists. Then, in view of Theorem~0.2~\cite{AP2020}, the function $\psi^+ (x,t) $ of (\ref{2.20})  solves  initial value problem for the free Dirac equation (\ref{DFree}). 
\qed

\bigskip

In the case of the mass  $m=0 ,\pm iH$ the Dirac equation obeys  Huygens' principle \cite{Wunsch,JPh_2021} and the 
function $ \psi^+ (x,t)$ is simplified as follows. First, we note that according to (3.13)~\cite{JPh_2021} 
\begin{eqnarray*}
K_1\left(r,t;- \frac{1}{2}H \right)
& =  &
K_1\left(r,t;  \frac{1}{2}H \right)=
\frac{1}{2}   e^{\frac{1}{2}H t}\,,\\
K_1\left(r,t; \frac{3}{2}H\right)
& =  & 
\frac{1}{4}   e^{-\frac{1}{2}  H t } \left(\left(1-H^2 r^2\right) e^{2 H t}+1\right)\,.
\end{eqnarray*}
Consequently,  by the definition (\ref{K1OPER}) of the operator $ {\cal K}_1$ we write
\begin{equation} 
{\cal K}_1\left(x,t,D_x;\frac{1}{2}H \right)[\varphi   (x ) ]
  =  
\label{K1minus12} 
  {\cal K}_1\left(x,t,D_x;- \frac{1}{2}H\right)[\varphi   (x ) ] = e^{\frac{1}{2}Ht}  V_{\varphi } (x, \phi (t) )\,,
\end{equation}
where $ \phi (t):=(1-e^{-Ht})/H$ and
\begin{eqnarray}
\label{K1plus32}
& &
 {\cal K}_1\left(x,t,D_x;\frac{3}{2}H\right)[\varphi   (x )] \\
& = & 
\frac{1}{2}   e^{ \frac{3}{2}  H t }  \left(1+ e^{-2 H t}\right) V_{\varphi _1 }(x,  \phi (t))   
-H^2 \frac{1}{2}  e^{ \frac{3}{2}  H t }  \phi (t)^2 V_{\varphi  }(x,  \phi (t))
+ H^2     e^{\frac{3}{2}  H t}  \int_{0}^{\phi (t) }    V_{\varphi   } (x,  s) 
  s    ds  \,. \nonumber  
\end{eqnarray}

\begin{corollary}
\mbox{\rm (i)} If $m=0$, then 
for the solution of the  Dirac equation  we obtain from (2.11)~\cite{JPh_2021}
 \[ 
\psi^+   (x,t) 
= 
  e^{-Ht}\left(  \partial_0 {\mathbb I}_4 +  e^{-Ht} \sum_{k=1,2,3} \gamma ^k \gamma ^0\partial_k- \frac{H}{2}{\mathbb I}_4   \right)  e^{Ht/2} \frac{\phi (t)}{4\pi }\int_{|y|=1}\psi^+ _0 (x+\phi (t)y)\,dS_y\,.
\]
\mbox{\rm (ii)} For $m=iH$ we have $M_+=  - \frac{1}{2}H  $ and $M_-=  \frac{3}{2}H$.
Then for the operators  $ {\cal K}_1\left(x,t,D_x;- \frac{1}{2}H\right)$ and $  {\cal K}_1\left(x,t,D_x;\frac{3}{2}H\right)$
we have representation (3.14)~\cite{JPh_2021} and (3.15)~\cite{JPh_2021}, respectively. 
Hence  with $\psi^+_0  (x) =(\psi^+ _{00}(x ) ,\psi^+ _{01}(x ) ,\psi^+ _{02}(x ) ,\psi^+ _{03}(x ) )^T $ 
\[
\psi^+   (x,t)
=
  e^{-Ht}\left(  \partial_0 {\mathbb I}_4 +  e^{-Ht}\sum_{k=1,2,3} \gamma ^k \gamma ^0\partial_k- \frac{H}{2}{\mathbb I}_4 +H \gamma ^0 \right)  \left (
   \begin{array}{cccc}
 {\cal K}_1(x,t,D_x;- \frac{1}{2}H)[\psi^+ _{00}(x ) ]  \\
{\cal K}_1(x,t,D_x;- \frac{1}{2}H)[\psi^+ _{01}(x )] \\
 {\cal K}_1(x,t,D_x;\frac{3}{2}H)[ \psi^+ _{02}(x )]  \\
 {\cal K}_1(x,t,D_x;\frac{3}{2}H)[ \psi^+ _{03}(x ) ]\\
   \end{array}
   \right ).
\]
Here ${\cal K}_1(x,t,D_x;- \frac{1}{2}H) $ and ${\cal K}_1(x,t,D_x;\frac{3}{2}H) $   are given by (\ref{K1minus12}) and (\ref{K1plus32}), respectively.\\
\mbox{\rm (iii)}   For the case of $m=-iH$ we have  $M_+  =  \frac{3}{2}H $,  $M_-=   - \frac{1}{2}H$,  
and the formula  is  
\[
\psi^+   (x,t) 
  = 
e^{-Ht}\left(  \partial_0+e^{-Ht}\sum_{k=1,2,3} \gamma ^k \gamma ^0\partial_k-\frac{H}{2}{\mathbb I}_4 -H\gamma ^0\right)\left (
   \begin{array}{cccc}
 {\cal K}_1(x,t,D_x;\frac{3}{2}H)[\psi^+ _{00}(x ) ]  \\
{\cal K}_1(x,t,D_x;\frac{3}{2}H)[ \psi^+ _{01}(x )] \\
 {\cal K}_1(x,t,D_x;- \frac{1}{2}H)[\psi^+ _{02}(x )]  \\
 {\cal K}_1(x,t,D_x;- \frac{1}{2}H)[\psi^+ _{03}(x ) ]\\
   \end{array}
   \right ) .
 \]
\end{corollary}

\section{Lochak-Majorana condition in de~Sitter spacetime}
\label{S2}
\setcounter{equation}{0}

Consider  the matrix-valued potential function 
\begin{equation}
\label{A}
A(x,t) = \alpha (x,t){\mathbb I}_4 +i\beta (x,t) \gamma ^5,\,\, \alpha ,\beta  \in {\mathbb R} , \quad \gamma ^5:=-i\gamma ^0\gamma ^1\gamma ^2\gamma ^3\,,  
\end{equation}
that takes values in the space 
$
{\mathcal M}= \Big\{\alpha {\mathbb I}_4  +i\beta \gamma ^5,\,\, \alpha ,\beta  \in {\mathbb R} \Big\}
$.  
Then we consider the Dirac equation  with the potential $ A(x,t)$, where $\alpha ,  \beta\in C^0([0,\infty);$ $ L^2({\mathbb R}^3)) $.
\begin{lemma}
\label{L2.1}
For the solution $\psi \in C^1([0,\infty); L^2({\mathbb R}^3))\cap C^0([0,\infty); H_{1}({\mathbb R}^3))$ of the  Dirac equation
\begin{equation}
\label{DE_LM}
  \dsp 
\left(  i {\gamma }^0    \partial_0   +i e^{-Ht}\sum_{k=1,2,3} {\gamma }^k  \partial_k  +i \frac{3}{2}    H {\gamma }^0     -m{\mathbb I}_4+ {\gamma }^0 V(x,t)\right)\psi=-A \psi  \,,
\end{equation}
with the  self-adjoint $V$ the following energy identity holds
\[
 \|\psi (x,t)\|_{L^2({\mathbb R}^3)}^2 
  =   
 e^{-3Ht} \|\psi (x,0)\|_{L^2({\mathbb R}^3)}^2  + 2\Im (m) e^{-3Ht} \int_0^t e^{3Hs}\int_{{\mathbb R}^3}  \psi^*(x,s) {\gamma }^0 \psi\,(x,s)\,  dx\, ds \,.
\]
\end{lemma}
\medskip

\ndt
{\bf Proof.} 
The equation (\ref{DE_LM}) can be written as follows
\[
{\mathscr{D}}_{dS}(x,t, \partial)\psi=  i{\gamma }^0A \psi  \,.
\]
Hence
\begin{eqnarray*}
  \dsp 
\frac{d}{dt} |\psi |^2 
& =  &
-3    H |\psi |^2- \Big(  e^{-Ht}\sum_{\ell =1,2,3}(\partial_\ell\psi)^* \,({\gamma }^\ell)^*{\gamma }^0  -i\ol{m}  \psi^*{\gamma }^0\Big) \psi-  i\psi ^* A^*{\gamma }^0     \psi \\
&  &
 -  \psi^* \left( e^{-Ht}\sum_{\ell =1,2,3} {\gamma }^0{\gamma }^\ell   \partial_\ell \psi        +im{\gamma }^0 \psi \right) +i  \psi^* {\gamma }^0A \psi \,.
\end{eqnarray*}
Now we simplify the terms with  the potential  $A$:
\[
 -  i \psi^* A^* {\gamma }^0     \psi 
+i \psi^*  {\gamma }^0A \psi =0\,.
\]
Thus,
\[
\frac{d}{dt} |\psi |^2 
  =  
-3    H |\psi |^2+2\Im (m)  \psi^*  {\gamma }^0 \psi-    e^{-Ht}\sum_{\ell =1,2,3}\partial_\ell(  \psi^* \,
 {\gamma }^0{\gamma }^\ell \psi  ) \,.  
\]
In view of the finite propagation speed property, it follows
\[
  \dsp 
\frac{d}{dt} \|\psi \|_{L^2({\mathbb R}^3)}^2 
  = 
-3    H \|\psi \|_{L^2({\mathbb R}^3)}^2+2\Im (m)\int_{{\mathbb R}^3}   \psi^*{\gamma }^0  \psi\, dx
\]
and, consequently, 
\[
  \dsp 
\frac{d}{dt} \left( e^{3Ht}\|\psi \|_{L^2({\mathbb R}^3)}^2  \right) 
  =  
2e^{3Ht}\Im (m)\int_{{\mathbb R}^3}  \psi^*{\gamma }^0 \psi\, dx\,.
\]
By integration we obtain 
\[
 e^{3Ht}\|\psi (x,t)\|_{L^2({\mathbb R}^3)}^2  
  =  
  \|\psi (x,0)\|_{L^2({\mathbb R}^3)}^2  + 2\Im (m)\int_0^t e^{3Hs}\int_{{\mathbb R}^3}  \psi^*(x,s) {\gamma }^0  \psi\,(x,s)\, dx \,ds \,.
\]
Lemma is proved. \qed

\begin{lemma}
\label{L3.2}
Assume that 
\begin{equation}
\label{Vgamma2}
V^T(x,t)\gamma ^2+\gamma ^2V (x,t)=0\,.
\end{equation}
For the solution $\psi \in C^1([0,\infty); L^2({\mathbb R}^3))\cap C^0([0,\infty); H_{1}({\mathbb R}^3))$ of the  Dirac 
equation (\ref{DE_LM}) one has
\[
 \int_{{\mathbb R}^3} \psi^T (x,t) \gamma ^2\psi (x,t)\,dx 
 = 
 e^{-3Ht}\left( \int_{{\mathbb R}^3} \psi^T (x,0) \gamma ^2\psi (x,0) \,dx\right)  \,.
\]
In particular, for $z \in {\mathbb C}$, $|z|=1$,  one has
\[
 \int_{{\mathbb R}^3} 2\Re \Big( \ol{z}\psi^T (x,t) \gamma ^2\psi (x,t)\Big) \,dx 
 = 
 e^{-3Ht} \int_{{\mathbb R}^3} 2\Re \Big(  \ol{z}\psi^T (x,0) \gamma ^2\psi (x,0) \Big) \,dx\,. 
\]
\end{lemma}
\medskip

\noindent
{\bf Proof.} We multiply the equation by $\psi^T\gamma ^2 $:
\begin{eqnarray*}
&  &
  \dsp 
 \psi^T\gamma ^2 \partial_0\psi   + e^{-Ht}\sum_{\ell =1,2,3}\psi^T\gamma ^2 {\gamma }^0{\gamma }^\ell  \partial_\ell\psi + \frac{3}{2}    H\psi^T\gamma ^2 \psi   +im\psi^T\gamma ^2{\gamma }^0  \psi-i\psi^T\gamma ^2V(x,t)\psi\\
&  &
=  A \psi^T\gamma ^2i{\gamma }^0\psi  \,.
\end{eqnarray*}
Here
\begin{eqnarray*}
&  &
\gamma ^2\gamma ^0 
= \left(
\begin{array}{cccc}
 {\mathbb O}_2 & -\sigma _2  \\
 -\sigma _2 & {\mathbb O}_2  \\
\end{array}
\right)\,,\quad \gamma ^0\gamma ^5=
\left(
\begin{array}{cccc}
 {\mathbb O}_2 & -{\mathbb I}_2  \\
 {\mathbb I}_2 & {\mathbb O}_2  \\
\end{array}
\right)=-\gamma ^5\gamma ^0,   \quad \gamma ^5\gamma ^2+\gamma ^2\gamma ^5=0\,,\\
&  &
\gamma ^2\gamma ^0\gamma ^1=
\left(
\begin{array}{cccc}
-i\sigma _3   & {\mathbb O}_2 \\
 {\mathbb O}_2 & i\sigma _3  \\
\end{array}
\right)
, \quad\gamma ^2\gamma ^0\gamma ^2=\gamma ^0,\quad \gamma ^2\gamma ^0\gamma ^3
=\left(
\begin{array}{cccc}
i\sigma _1   & {\mathbb O}_2 \\
 {\mathbb O}_2 & -i\sigma _1  \\
\end{array}
\right)\,.
\end{eqnarray*}
Furthermore,  
\[
  \dsp 
\partial_0\psi  
  =  
- e^{-Ht}\sum_{\ell =1,2,3}{\gamma }^0{\gamma }^\ell \partial_\ell\psi  -   \frac{3}{2}    H\psi      -im{\gamma }^0  \psi+iV(x,t) \psi+  i{\gamma }^0A \psi  \,, 
\]
and due to (\ref{Vgamma2}) we obtain
\begin{eqnarray*}
 \partial_t  (\psi^T\gamma ^2\psi)
 & = &
\Bigg(- \sum_{\ell =1,2,3} e^{-Ht}(\partial_\ell\psi)^T ({\gamma }^\ell)^T{\gamma }^0  - \frac{3}{2}    H\psi^T     -im \psi^T {\gamma }^0+  i\psi^T  A{\gamma }^0  \Bigg)\gamma ^2 \psi\\
&  &
+ \psi^T\gamma ^2   \Bigg(- e^{-Ht}\sum_{\ell =1,2,3}{\gamma }^0{\gamma }^\ell \partial_\ell\psi  -   \frac{3}{2}    H\psi      -im{\gamma }^0  \psi+  i{\gamma }^0 A\psi \Bigg)\,.
\end{eqnarray*}
Consider  the terms with potential:
\begin{eqnarray*}
\Big( i\psi^T  A{\gamma }^0  \Big)\gamma ^2 \psi
+ \psi^T\gamma ^2   \Big( i{\gamma }^0A \psi \Big)
 & = &
i\psi^T  \Bigg(\alpha {\mathbb I}_4 +i\beta \gamma ^5\Bigg){\gamma }^0 \gamma ^2 \psi
+ \psi^T\gamma ^2    i{\gamma }^0\Bigg(\alpha {\mathbb I}_4 +i\beta \gamma ^5 \Bigg) \psi\\
 & = &
i\psi^T  \Bigg(\alpha{\mathbb I}_4 {\gamma }^0\gamma ^2+i\beta \gamma ^5{\gamma }^0 \gamma ^2
+     \alpha {\mathbb I}_4\gamma ^2 {\gamma }^0 +i\beta  \gamma ^2{\gamma }^0 \gamma ^5  \Bigg)\psi\\
 & = &
i\psi^T  \Big(i\beta \gamma ^5{\gamma }^0 \gamma ^2
+i\beta  \gamma ^2 {\gamma }^0\gamma ^5  \Big)\psi=0\,.
\end{eqnarray*}
Thus,
\[ 
\partial_t  (\psi^T\gamma ^2\psi)
  =  
 - 3    H\psi^T \gamma ^2 \psi- e^{-Ht}\sum_{\ell =1,2,3} \Big\{ (\partial_\ell\psi)^T ({\gamma }^\ell)^T{\gamma }^0 \gamma ^2 \psi+\psi^T\gamma ^2 {\gamma }^0{\gamma }^\ell \partial_\ell\psi \Big\} \,.
\]
For the sum in the last equation, we have
\begin{eqnarray*}
({\gamma }^1)^T{\gamma }^0 \gamma ^2
= \gamma ^2{\gamma }^0{\gamma }^1\,,\quad
({\gamma }^2)^T{\gamma }^0 \gamma ^2
= \gamma ^2{\gamma }^0  \gamma ^2\,,\quad
({\gamma }^3)^T{\gamma }^0 \gamma ^2
= \gamma ^2{\gamma }^0{\gamma }^3\,.
\end{eqnarray*}
It follows
\[
 \partial_t  (\psi^T\gamma ^2\psi)
=
 - 3    H\psi^T \gamma ^2 \psi- e^{-Ht}\sum_{\ell =1,2,3}  \partial_\ell\Big( \psi^T\gamma ^2 {\gamma }^0{\gamma }^\ell \psi \Big)
\]
and, consequently, 
\[
 \partial_t \int_{{\mathbb R}^3} (\psi^T\gamma ^2\psi)\,dx
 =
 - 3    H\int_{{\mathbb R}^3} \psi^T \gamma ^2 \psi\,dx\,.
\]
Thus, the first statement of the lemma is proved. To prove the second statement, we use 
$\ol{\gamma _2}=  -\gamma _2$ and  recall the formula from  \cite{Bachelot}:
\begin{equation}
\label{3.17}
|\psi -z\gamma ^2 \ol{\psi }|^2= 2|\psi |^2 +2 \Re (\ol{z} \psi^T \gamma ^2 \psi) \,,
\end{equation}
where $z \in {\mathbb C}$,   $|z|=1$. 
The lemma is proved. \qed

\begin{lemma}
For the solution $\psi \in C^1([0,\infty); L^2({\mathbb R}^3))\cap C^0([0,\infty); H_{(1)}({\mathbb R}^3))$ of the  Dirac 
equation (\ref{DE_LM}) one has
\begin{eqnarray*}
\int_{{\mathbb R}^3} |\psi (x,t) -z\gamma ^2 \ol{\psi  (x,t)}|^2\,dx 
& =  &
 e^{-3Ht} \Bigg\{\int_{{\mathbb R}^3} |\psi (x,0) -z\gamma ^2 \ol{\psi  (x,0)}|^2\,dx  \\
&  &
+ 4\Im (m)\int_0^t e^{3Hs}\int_{{\mathbb R}^3}   \psi^*(x,s) {\gamma }^0  \psi\,(x,s)\, dx \,ds \Bigg\} \,.
\end{eqnarray*}
\end{lemma}
\medskip

\noindent
{\bf Proof.} From Lemma~\ref{L2.1}, Lemma~\ref{L3.2}, and (\ref{3.17})
\begin{eqnarray*}
&  &
\int_{{\mathbb R}^3} |\psi (x,t) -z\gamma ^2 \ol{\psi  (x,t)}|^2\,dx \\
& =  &
\int_{{\mathbb R}^3} \Big(2|\psi |^2 +2 \Re (\ol{z} \psi^T \gamma ^2 \psi)\Big)\,dx \\
& =  &
2 \left(  e^{-3Ht} \|\psi (x,0)\|^2  + 2\Im (m) e^{-3Ht} \int_0^t e^{3Hs}\int_{{\mathbb R}^3}  \psi^*(x,s) {\gamma }^0  \psi\,(x,s) dx ds\right) \\
&  &
+  2 \int_{{\mathbb R}^3}\Re \Big ( \ol{z} \psi^T (x,0)\gamma ^2 \psi(x,0) \Big )\,dx \,.
\end{eqnarray*}
That is,
\begin{eqnarray*}
\int_{{\mathbb R}^3} |\psi (x,t) -z\gamma ^2 \ol{\psi  (x,t)}|^2\,dx 
& =  &
2 e^{-3Ht} \left(  \|\psi (x,0)\|^2 + \int_{{\mathbb R}^3}  \Re \Big(\ol{z}  \psi^T (x,0) \gamma ^2\psi (x,0) \Big) \,dx\right) \\
&  &
+ 4\Im (m) e^{-3Ht} \int_0^t e^{3Hs}\int_{{\mathbb R}^3}  \psi^*(x,s) {\gamma }^0  \psi\,(x,s)\, dx \,ds  \,.
\end{eqnarray*}
Lemma is proved. \qed
\smallskip

The following identity is easily seen for the function defined in (\ref{LMC}):
\[
\rho ^2 (\psi )
 = 
  | \psi ^*\gamma ^0\psi|^2 +  | \psi ^*\gamma ^0\gamma ^5\psi|^2 = (|\psi _1|^2+|\psi _2|^2-|\psi _3|^2-|\psi _4|^2)^2+ (2\Im (\psi _1\ol{\psi _3})+2\Im (\psi _2\ol{\psi _4}))^2\,.
\]
\begin{corollary}
\label{C2.5}
Assume that (\ref{Vgamma2}) is fulfilled. 
(i) If $\Im (m)=0$, then
\[
\int_{{\mathbb R}^3} |\psi (x,t) -z\gamma ^2 \ol{\psi  (x,t)}|^2\,dx 
 = 
e^{-3Ht} \int_{{\mathbb R}^3} |\psi (x,0) -z\gamma ^2 \ol{\psi  (x,0)}|^2\,dx  \,.
\]
(ii) If $\psi (x,0) -z\gamma ^2 \ol{\psi  (x,0)}=0$ and $\Im (m) \not=0$, then
\[
\int_{{\mathbb R}^3} |\psi (x,t) -z\gamma ^2 \ol{\psi  (x,t)}|^2\,dx 
 =  4\Im (m) e^{-3Ht} \int_0^t e^{3Hs}\int_{{\mathbb R}^3}  \overline{\psi^T(x,s) }{\gamma }^0\cdot \psi\,(x,s)\, dx \,ds  \,.
\]
(iii) if $\psi (x,0) -z\gamma ^2 \ol{\psi  (x,0)}=0$, then
\[
\int_{{\mathbb R}^3} |\psi (x,t) -z\gamma ^2 \ol{\psi  (x,t)}|^2\,dx 
 \leq 
4|\Im (m)| e^{-3Ht} \int_0^t e^{3Hs}\int_{{\mathbb R}^3}  \rho (x,s)\, dx \,ds \,. 
\]
\end{corollary}

Note, the statements (i),(iii)   of  Proposition~I.1~\cite{Bachelot} imply that  for  $z \in {\mathbb C}$, $|z|=1$, the condition  $\psi -z\gamma ^2 \ol{\psi }=0$    is equivalent to $\rho ^2 (\psi )=0$.

\section{Proof of Theorem~\ref{T4.1}}
\label{S3}
\setcounter{equation}{0}

In order to prove the existence of the global solution to  the Cauchy problem for the semilinear Dirac equation in the Minkowski space,  Bachelot~\cite{Bachelot} used the successive approximations and   appealed to the estimates, which are obtained by the replacing the generators of the Poincar\'e  group with the Fermi operators. We use the energy estimate  obtained in subsection~\ref{SSEE} that 
allows   us  to expand  the result from \cite{Bachelot} to the Dirac equation in the de~Sitter spacetime.

\medskip

  Let $\Psi $ be the solution of 
\begin{equation}
\label{B31}  
\cases{ \dsp  {\mathscr{D}}_{dS}(x,t, \partial)\Psi (x,t)  =0  \,,\cr
\Psi (x,0)=\Psi _0 (x)\,.
}
\end{equation} 
The finite propagation speed property for the equation in the de~Sitter spacetime implies that the support of $\Psi =\Psi (x,t) \in C^\infty({\mathbb R}^3\times {\mathbb R}_+  ; {\mathbb C}^4)$ is in the same compact subset 
of ${\mathbb R}^3$ for all $t>0$. Then according to Lemma~\ref{L1.2} 
\begin{equation}
\label{4.26}
 e^{  -\frac{1}{2}\delta_+ t}\|\Psi (t) \|_k 
  \leq   
c   \|\Psi _0  \|_k \quad \mbox{\rm for all}\quad t>0\,.
\end{equation}
 We look for the solution of (\ref{NDE_CP}) in the form
\begin{equation}
\label{B33}
\psi =\Psi +\chi \,,
\end{equation}
where $\Psi  $ solves (\ref{B31}). 

Consider the nonlinear term. 
First of all, we note that, according to Corollary~\ref{C2.5}, (\ref{Falphabeta}),  and (\ref{LMC}), 
if $\Im (m)=0$, then 
\[
\dsp F\left( \Psi^*   (x,t) \gamma ^0 \Psi   (x,t) \,,\,\Psi^*  (x,t)\gamma ^0\gamma ^5 \Psi    (x,t) \right) 
 = 
0 \quad \mbox{\rm for all}\quad t \geq 0,\quad x \in {\mathbb R}^3 \, .
\]
Further, (\ref{F}) 
can be written as follows
\[
F\left(\xi ,\eta  \right)= \alpha \left(\xi ,\eta  \right){\mathbb I}_4 +i\beta \left(\xi ,\eta  \right) \gamma ^5,\quad \xi = \psi^*\gamma ^0 \psi \in {\mathbb R},\quad
\eta =\psi^*\gamma ^0\gamma ^5 \psi \in {\mathbb R}\,.
\]
It is evident that with the solution $\Psi   (x,t)  $ we can write
\begin{eqnarray*}
|(\Psi (x,t)+\chi (x,t))^*\gamma ^0 (\Psi(x,t) +\chi (x,t))- \Psi ^* (x,t)  \gamma ^0  \Psi(x,t)| 
& \!\! \leq \!\!  &
C \big(|\chi (x,t)|+|\Psi  (x,t)|\big)^2,\\
|(\Psi (x,t)+\chi (x,t)) ^*\gamma ^0\gamma ^5 (\Psi (x,t)+\chi (x,t))-  (\Psi ^*(x,t)\gamma ^0\gamma ^5  \Psi  (x,t) |
& \!\! \leq  \!\! &
C \big(|\chi (x,t)|+|\Psi  (x,t)|\big)^2 
\end{eqnarray*}
and, consequently, 
\begin{eqnarray*}
 \alpha \left( \psi^*\gamma ^0 \psi\,,\,\psi^*\gamma ^0\gamma ^5 \psi \right)\psi 
& = &
  \alpha \left( (\Psi +\chi )^*\gamma ^0 (\Psi +\chi )\,,\,(\Psi +\chi )^*\gamma ^0\gamma ^5 (\Psi +\chi ) \right)\psi \\
& = &
  \alpha \left( \Psi^*\gamma ^0 \Psi  \,,\,\Psi^*\gamma ^0\gamma ^5 \Psi  \right)\psi +\alpha_1 \left(\chi ,\Psi \right)  \\
& = &
\alpha_1 \left(\chi ,\Psi \right)\,,\\
 \beta  \left(\psi^*\gamma ^0 \psi\,,\,\psi^*\gamma ^0\gamma ^5 \psi \right)\psi 
& = &
  \beta \left((\Psi +\chi )^*\gamma ^0 (\Psi +\chi )\,,\,(\Psi +\chi )^*\gamma ^0\gamma ^5 (\Psi +\chi ) \right)\psi \\
& = &
  \beta  \left(\Psi^*\gamma ^0 \Psi  \,,\,\Psi^*\gamma ^0\gamma ^5 \Psi  \right)\psi +\beta _1 \left(\chi ,\Psi \right)  \\
& = &
\beta _1 \left(\chi ,\Psi \right)\,,
\end{eqnarray*}
where  $ \alpha_1 ,\beta _1  \in C^\infty({\mathbb C}^8; {\mathbb C}^4)$, and, as the functions of $\chi  $,
\begin{eqnarray}
\label{B35a}
|\alpha _1\left(\chi \,,\,\Psi  (x,t)\right)| 
& = &
O\Big( |\chi |\big(|\chi |+|\Psi (x,t) |\big)^2\Big)\quad \mbox{\rm as} \quad |\chi| \to 0\,, \\
\label{B35b}
|\beta  _1\left(\chi \,,\,\Psi  (x,t)\right)| 
& = &
O\Big( |\chi |\big(|\chi |+|\Psi (x,t) |\big)^2\Big)\quad \mbox{\rm as} \quad |\chi| \to 0\,.
\end{eqnarray}

Further,
the Cauchy problem becomes 
\begin{equation}
\label{B34}
\cases{ \dsp  i {\gamma }^0   {\mathscr{D}}_{dS}(x,t, \partial)\chi   =f_1\left(\chi \,,\,\Psi  \right)   \,,\cr
\chi  (x,0)=\varepsilon \chi _0 (x)\,,
}
\end{equation}
where, in view of (\ref{B31}), we have denoted
\begin{eqnarray*}
f_1\left(\chi \,,\,\Psi  \right)
& := &
  F\left((\Psi^*  +\chi^*)\gamma ^0 (\Psi  +\chi  )\,,\,(\Psi^*  +\chi^*)\gamma ^0\gamma ^5 (\Psi  +\chi  ) \right)  \Psi  \\
&   &
 + F\left((\Psi^*  +\chi^*)\gamma ^0 (\Psi  +\chi  )\,,\,(\Psi^*  +\chi^*)\gamma ^0\gamma ^5 (\Psi  +\chi  ) \right)  \chi    
\end{eqnarray*}
that can be rewritten  similar to (\ref{A}) as $ f_1\left(\chi \,,\,\Psi  \right)=-A \chi $. 
Hence,  $ f_1  \in C^\infty({\mathbb C}^8; {\mathbb C}^4)$, while (\ref{B35a}) and (\ref{B35b}) imply
\[
|f_1\left(\chi \,,\,\Psi  (x,t)\right)| =O\Big( |\chi |\big(|\chi |+|\Psi (x,t) |\big)^2\Big)\quad \mbox{\rm as} \quad |\chi| \to 0\,.
\]
We look for the function $\chi  $ as a limit of the sequence  $\{\chi ^{(k)}\}_1^\infty $ that is defined  by
\[ 
\cases{ \dsp  i {\gamma }^0    {\mathscr{D}}_{dS}(x,t, \partial)\chi^{(k)}   =f_1\left(\chi^{(k-1)} \,,\,\Psi  \right)   \,,\cr
\chi ^{(k)} (x,0)=\varepsilon \chi _0 (x), \quad k=1,2,\ldots\,,
}
\]
and $\chi ^{(0)}(x,t)\equiv 0 $. The finite propagation speed property implies that the supports of the functions $\chi^{(k)} =\chi^{(k)} (x,t) $ are in the same compact subset 
of ${\mathbb R}^3$ for all $t>0$ and $k \geq 0$. Lemmas~\ref{L1.2},\ref{L2.1}  and the estimate
\[
 \|\chi ^{(1)}(t)\|_s 
\leq 
 c e^{\frac{1}{2}\delta _+ t}\| \chi ^{(1)} (0) \|_{s}+ c e^{\frac{1}{2}\delta _+ t} \int_0^t  e^{-\frac{1}{2}\delta _+ \tau }  \| f_1\left(\chi^{(0)}(\tau ) \,,\,\Psi (\tau ) \right)  \|_{s} \,d\tau   \quad \mbox{\rm for all}\quad t>0\,,
\]
imply
\[
 \|\chi ^{(1)}(t)\|_s  
=
c \varepsilon  e^{\frac{1}{2}\delta _+t} \|\chi _0 \|_s    \quad \mbox{\rm for all}\quad t>0\,.
\]
Corollary~6.4.5~\cite{Hor} and Lemma~\ref{L1.2} imply for $k=2,3,\ldots $ 
the estimate
\begin{eqnarray}
\label{3.8}
\hspace{-0.3cm} &  &
\| \chi^{(k)}  (t) \|_{s}  \nonumber  \\
\hspace{-0.3cm} & \leq  &
e^{\frac{1}{2}\delta _+ t} \Bigg( c\varepsilon \| \chi _0 (x) \|_{s}+ 2c \int_0^t  e^{-\frac{1}{2}\delta _+\tau }  \| f_1\left(\chi^{(k-1)} \,,\,\Psi  \right) (\tau ) \|_{s} \,d\tau  \Bigg) \nonumber \\
\hspace{-0.3cm} & \leq  &
e^{\frac{1}{2}\delta _+ t} \Bigg( c\varepsilon \| \chi _0 (x) \|_{s}+ 2c \int_0^t  e^{-\frac{1}{2}\delta _+\tau }  \| \chi^{(k-1)} (\tau )\|_s \Big( | \chi^{(k-1)} (\tau )|_{\left[\frac{s}{2}\right]}
+ |\Psi  (\tau )|_{\left[\frac{s}{2}\right]} \Big)^2   \,d\tau  \Bigg),
\end{eqnarray}
where $ \delta_+ = -3H +2|\Im (m)|<0$ and $\left[\frac{s}{2}\right] $ is the integer part of $\frac{s}{2} $, while 
\[
|\Psi  (t )|_{s}:=\sup_{|\alpha| \leq s}  \|\partial_x^\alpha \Psi  (x, t )\|_{L^\infty({\mathbb R}^3)}.
\] 
Now we apply Sobolev embedding theorem and Lemma~\ref{L1.2}  to the function $ \Psi $:
\[
 | \Psi  (t)  |_{\left[\frac{s}{2}\right]} \leq \| \Psi  (t) \|_{s} 
 \leq 
 e^{\frac{1}{2}\delta _+ t}\| \Psi_0  \|_{s}\quad \mbox{\rm for all}\quad t>0\,.
\]
For the given $s$ and $n$ we define 
\[
a_n(t):=\sup_{{0\leq \tau \leq t,\,\,0\leq k\leq n}} e^{ -\frac{1}{2}\delta _+ \tau }\| \chi^{(k)}  (\tau ) \|_{s}\,,\qquad A_n:=\sup_{t \in {\mathbb R}_+} a_n(t)\,.
\]
Then  for $s \geq 6$ by (\ref{3.8}) we derive 
\[
\| \chi^{(k)}  (t) \|_{s}  \\
\leq 
e^{\frac{1}{2}\delta _+ t} \Bigg( c\varepsilon \| \chi _0 \|_{s}+ 2 c\int_0^t  e^{\frac{1}{2}\delta _+ \tau }  \| \chi^{(k-1)} (\tau )\|_s \Big( e^{-\frac{1}{2}\delta _+ \tau }\| \chi^{(k-1)} (\tau )\|_{s}
+ e^{-\frac{1}{2}\delta _+\tau }\|\Psi  (\tau )\|_{s} \Big)^2   \,d\tau  \Bigg)
\]
and
\begin{eqnarray*} 
a_n(t) 
& \leq  &
c\varepsilon \| \chi _0 \|_{s}+ 2 c\int_0^t  e^{\delta _+ \tau }\Big( e^{ -\frac{1}{2}\delta _+ \tau }  \| \chi^{(n-1)} (\tau )\|_s \Big) \Big( e^{-\frac{1}{2}\delta _+ \tau }\| \chi^{(n-1)} (\tau )\|_{s}
+ e^{-\frac{1}{2}\delta _+\tau }\|\Psi  (\tau )\|_{s} \Big)^2   \,d\tau \\
& \leq  &
c\varepsilon \| \chi _0\|_{s}+ 2 c\int_0^t  e^{\delta _+ \tau }\Big( e^{-\frac{1}{2}\delta _+ \tau }  \| \chi^{(n-1)} (\tau )\|_s \Big) \Big( e^{-\frac{1}{2}\delta _+ \tau }\| \chi^{(n-1)} (\tau )\|_{s}
+ \| \Psi_0  \|_{s} \Big)^2   \,d\tau 
\end{eqnarray*}
that implies (with the new $c$ depending on $\| \Psi_0  \|_{s} $) for $n=1,2,\ldots$ 
\begin{eqnarray} 
\label{4.30}
a_{n}(t) 
& \leq  &
c\varepsilon \| \chi _0  \|_{s}+ 2c\big( 1+ A_{n-1}\big)^2 \int_0^t  e^{\delta _+  \tau } a_{n-1}(\tau )   \,d\tau  \nonumber  \\
& \leq  &
c\varepsilon \| \chi _0  \|_{s}+ 2c\big( 1+ A_{n-1}\big)^2 \int_0^t  e^{\delta _+  \tau } a_{n }(\tau )   \,d\tau  \,.
\end{eqnarray}
Denote
\[
y(t):=\int_0^t  e^{\delta _+  \tau } a_{n}(\tau )   \,d\tau \quad \mbox{\rm then} \quad  y'(t)=  e^{\delta _+  t } a_{n }(t)
\]
and according to (\ref{4.30}) we obtain
\[ 
y'(t) 
\leq  
e^{\delta _+  t }c\varepsilon \| \chi _0\|_{s}+ 2c\big( 1+ A_{n-1}\big)^2 e^{\delta _+  t }y(t)\,.  
\]
It follows
\[
\frac{d}{d t}\left(  e^{-2c\big( 1+ A_{n-1}\big)^2 \frac{1}{\delta _+  } \left( e^{\delta _+  t  }-1 \right)}  y(t)\right)
  \leq  
  e^{-2c\big( 1+ A_{n-1}\big)^2 \frac{1}{\delta _+  } \left( e^{\delta _+  t  }-1 \right)}e^{\delta _+  t }c\varepsilon \| \chi _0  \|_{s}  
\]
and, consequently,
\[ 
  e^{-2c\big( 1+ A_{n-1}\big)^2 \frac{1}{\delta _+  } \left( e^{\delta _+  t  }-1 \right)} y(t)
\leq 
c\varepsilon \| \chi _0 \|_{s} \int_0^t   e^{-2c\big( 1+ A_{n-1}\big)^2 \frac{1}{\delta _+} \left(e^{\delta _+ \tau   }   -1\right)+\delta _+ \tau  } \, d\tau\,,  
\]
since $y(0)=0$. Hence
\begin{eqnarray*} 
 y(t)
& \leq  &
 e^{-2c\big( 1+ A_{n-1}\big)^2 \frac{1}{\delta _+ } \left(1 - e^{\delta _+ t  }\right)} 
c\varepsilon \| \chi _0 \|_{s} \int_0^t   e^{-2c\big( 1+ A_{n-1}\big)^2 \frac{1}{\delta _+ } \left( e^{\delta _+ \tau   }-1 \right)+\delta _+ \tau  } \, d\tau  \\
& \leq  &
 e^{2c \big( 1+ A_{n-1}\big)^2 \frac{1}{\delta _+ }   e^{\delta _+ t  } } 
c\varepsilon \| \chi _0 \|_{s} \int_0^t   e^{-2c\big( 1+ A_{n-1}\big)^2 \frac{1}{\delta _+ }  e^{\delta _+ \tau   }+\delta _+ \tau  } \, d\tau\,.
\end{eqnarray*}
Further, in view of (\ref{4.30}) we obtain
\begin{eqnarray*}  
a_{n}(t) 
& \leq  &
c\varepsilon \| \chi _0  \|_{s}+ 2c\big( 1+ A_{n-1}\big)^2 \int_0^t  e^{\delta _+  \tau } a_{n }(\tau )   \,d\tau  \\
& \leq  &
c\varepsilon \| \chi _0  \|_{s}+ 2c\big( 1+ A_{n-1}\big)^2 y(t) \\
& \leq  &
c\varepsilon \| \chi _0  \|_{s}+ 2c\big( 1+ A_{n-1}\big)^2
 e^{ 2c \big( 1+ A_{n-1}\big)^2 \frac{1}{\delta _+ }   e^{\delta _+ t  } } 
c\varepsilon \| \chi _0 \|_{s} \int_0^t   e^{-2c\big( 1+ A_{n-1}\big)^2 \frac{1}{\delta _+ }  e^{\delta _+ \tau   }+\delta _+ \tau  } \, d\tau \\
& \leq  &
c\varepsilon \| \chi _0 \|_{s}\left\{1 + 2c\big( 1+ A_{n-1}\big)^2
 e^{ 2c \big( 1+ A_{n-1}\big)^2 \frac{1}{\delta _+ }   e^{\delta _+ t  } } 
 \int_0^t   e^{-2c\big( 1+ A_{n-1}\big)^2 \frac{1}{\delta _+ }  e^{\delta _+ \tau   }+\delta _+ \tau  } \, d\tau \right\}\,.
\end{eqnarray*}
It follows
\[ 
A_{n} 
\leq 
c\varepsilon \| \chi _0 \|_{s}\left\{1 + 2c\big( 1+ A_{n-1}\big)^2
 e^{ 2c \big( 1+ A_{n-1}\big)^2 \frac{1}{\delta _+ }   e^{\delta _+ t  } } 
 \int_0^t   e^{-2c\big( 1+ A_{n-1}\big)^2 \frac{1}{\delta _+ }  e^{\delta _+ \tau   }+\delta _+ \tau  } \, d\tau \right\}\,.
\]
On the other hand,
\[  
\int_0^t   e^{-2c\big( 1+ A_{n-1}\big)^2 \frac{1}{\delta _+ }  e^{\delta _+ \tau   }+\delta _+ \tau  } \, d\tau 
  \leq  
\frac{1}{  {2c\big( 1+ A_{n-1}\big)^2}} e^{\frac{ -2c( 1+ A_{n-1})^2}{\delta_+} }
\] 
leads to
\begin{eqnarray*} 
A_{n} 
& \leq &
c\varepsilon \| \chi _0 \|_{s}\left\{1 + 
 e^{ 2c \big( 1+ A_{n-1}\big)^2 \frac{1}{\delta _+ }   e^{\delta _+ t  }  -\frac{2c( 1+ A_{n-1})^2}{\delta_+} }\right\}  
  \leq  
c\varepsilon \| \chi _0 \|_{s}\left\{1 + 
 e^{  -\frac{2c( 1+ A_{n-1})^2}{\delta_+} }\right\} \,.
\end{eqnarray*} 
Finally,
\begin{eqnarray} 
A_{n} 
& \leq &
2c\varepsilon \| \chi _0 \|_{s} 
 \exp \left\{   -\frac{2c }{\delta_+} \right\}  \exp \left\{  -\frac{4c A_{n-1}(1+ A_{n-1} )}{\delta_+}\right\}\nonumber \\
\label{4.32}
& \leq &
C\varepsilon  \exp \left\{  -\frac{4c }{\delta_+}A_{n-1}(1+ A_{n-1} )\right\}\,, \quad C:=2c \| \chi _0 \|_{s}\exp \left\{   -\frac{2c }{\delta_+} \right\}\,.
\end{eqnarray} 
Let $ \varepsilon _0$ be such that 
\[
2 \varepsilon_0 C<1 \quad \mbox{\rm and}\quad 
 -\frac{16c }{\delta_+}2C\varepsilon < \ln 2 \,.
\]
If
\[
A_{n-1}\leq 2C\varepsilon  \quad \mbox{\rm and} \quad\varepsilon \leq \varepsilon _0 \,, 
\]
then due to (\ref{4.32})  we obtain
\begin{eqnarray*} 
A_{n} 
& \leq  &
C\varepsilon  \exp \left\{  -\frac{4c }{\delta_+}A_{n-1}(1+ A_{n-1} )\right\}  \\
& \leq  &
C\varepsilon  \exp \left\{  -\frac{4c }{\delta_+}2C\varepsilon(1+ 2C\varepsilon)\right\}  \\
& \leq  &
C\varepsilon  \exp \left\{  -\frac{16c }{\delta_+}2C\varepsilon\right\}  \\
& \leq  &
 2 C\varepsilon\,.
\end{eqnarray*}
Thus, for given $s$ and for all $n \geq 1$ we have proved the estimate
\[
\sup_{t \in {\mathbb R}_+}\,\, \sup_{{0\leq \tau \leq t\,,\,0\leq k\leq n}} e^{-\frac{1}{2} \delta_+ \tau }\| \chi^{(k)}  (\tau ) \|_{s}\leq 
4c \| \chi _0 \|_{s}\exp \left\{   -\frac{2c }{\delta_+} \right\}\varepsilon \quad \mbox{\rm  for all} \quad n \geq 1\,.
\]
The last estimate, (\ref{4.26}), and Sobolev inequality imply
\begin{equation} 
\label{B47}
\sup_{ n=0,1,2\ldots } \,\,\sup_{x \in {\mathbb R}^3 ,\,\,t \in {\mathbb R}_+}\left\{e^{-\frac{1}{2} \delta_+ t } | \chi^{(n)}  (x,t) |, e^{-\frac{1}{2} \delta_+ t } | \Psi (x,t) |   \right\} = r < \infty\,.
\end{equation} 
Hence, for all $(x,t) \in {\mathbb R}^3\times {\mathbb R}_+$ we have
\begin{eqnarray}
\label{4.29} 
&  &
\Big|f_1(\chi^{(k-1)} (x,t); \Psi (x,t) ) -f_1(\chi^{(k-2)} (x,t); \Psi (x,t) )  \Big| \\
& \leq &
\Big|  \chi^{(k-1)} (x,t)  - \chi^{(k-2)} (x,t)   \Big| \sup_{{\xi , \eta  \in {\mathbb C}\,\,| \xi |, | \eta | \leq r}}   | \nabla_{\xi ,\eta}f_1 (\xi ,\eta ) |\,.\nonumber
\end{eqnarray}
Consider 
\[
\dsp  i {\gamma }^0    {\mathscr{D}}_{dS}(x,t, \partial) \Big( \chi^{(k)} -\chi^{(k-1)} \Big)=
f_1\left(\chi^{(k-1)} \,,\,\Psi  \right) - f_1\left(\chi^{(k-2)} \,,\,\Psi  \right)  \,,\quad k=1,2,\ldots\,.
\]
By Lemma~\ref{L1.2}
for the solution of the last equation considering  (\ref{4.29}) and the initial values, 
one has 
\begin{eqnarray*}
&  &
e^{-\frac{1}{2} \delta_+ t}\|  \chi^{(k)} (t)-\chi^{(k-1)}(t) \|_{L^2({\mathbb R}^3)} \\
& \leq  &
c\int_0^t  e^{-\frac{1}{2} \delta_+ s}  \|f_1\left(\chi^{(k-1)}(x, \tau ) \,,\,\Psi (x, \tau )  \right) - f_1\left(\chi^{(k-2)}(x, \tau )  \,,\,\Psi (x, \tau )  \right) \|_{L^2({\mathbb R}^3)} \,ds\\ 
& \leq  &
c \Bigg( \sup_{{\xi , \eta  \in {\mathbb C},\,\| \xi |, | \eta | \leq r}}   |\nabla_{\xi ,\eta}f_1 (\xi ,\eta ) | \Bigg)\int_0^t  e^{-\frac{1}{2} \delta_+ s}  \|
 \chi^{(k-1)} (x,t)  - \chi^{(k-2)} (x,t)   \|_{L^2({\mathbb R}^3)} \,ds\,.
\end{eqnarray*}
It follows
\begin{eqnarray*}
&  &
e^{-\frac{1}{2} \delta_+ t}\|  \chi^{(k)} (t)-\chi^{(k-1)}(t) \|_{L^2({\mathbb R}^3)} \\
& \leq  &
c^2 \Bigg( \sup_{{\xi , \eta  \in {\mathbb C},\,\,| \xi |, | \eta | \leq r}}   |\nabla_{\xi ,\eta}f_1 (\xi ,\eta ) | \Bigg)^2 \int_0^t \,ds \int_0^{s} e^{-\frac{1}{2} \delta_+ s_1}  \|
 \chi^{(k-2)} (x,s_1)  - \chi^{(k-3)} (x,s_1)   \|_{L^2({\mathbb R}^3)} \,ds_1 
\end{eqnarray*}
and, consequently, 
\[
e^{-\frac{1}{2} \delta_+ t}\|  \chi^{(k)} (t)-\chi^{(k-1)}(t) \|_{L^2({\mathbb R}^3)} 
\leq 
C\| \chi _0 (x) \|_{L^2({\mathbb R}^3)}
\frac{(Ct)^k}{k!}, \quad \mbox{\rm for all} \quad k=1,2,\ldots\,. 
\]
Hence, the sequence $ e^{-\frac{1}{2} \delta_+ t}  \chi^{(k)} (t)$  converges  to some  $ e^{-\frac{1}{2} \delta_+ t} \chi \in C^0([0,\infty); (L^2({\mathbb R}^3)^4)$, that is,
\[
\lim_{k \to \infty} e^{-\frac{1}{2} \delta_+ t}  \chi^{(k)} (t) = e^{-\frac{1}{2} \delta_+ t}  \chi  (t) 
\]
uniformly on every compact subset of ${\mathbb R}^3 $. By (\ref{B47})
\[
\lim_{k \to \infty} f_1\left(\chi^{(k)} \,,\,\Psi  \right) = f_1\left(\chi \,,\,\Psi  \right)  \quad \mbox{\rm in} \quad C^0([0,\infty); (L^2({\mathbb R}^3)^4)\,.
\]
Thus, $\chi  $ solves (\ref{B34}), while $ \psi $ solves  (\ref{NDE_CP})  and
\begin{equation} 
\label{B48}
\sup_{t \in {\mathbb R}_+ }e^{-\frac{1}{2} \delta_+ t} \| \psi   (t) \|_{s}  < \infty \quad \mbox{\rm for } \quad s \geq 6
\end{equation} 
implies 
\[
\sup_{t \in {\mathbb R}_+ }e^{-\frac{1}{2} \delta_+ t} | \psi   (x,t)|_{s^\prime }  < \infty\quad \mbox{\rm for } \quad s^\prime  \leq s-1 \,.
\]
By Gagliardo-Nirenberg inequality for any integer $s \geq 6$ we have
\begin{equation} 
\label{B49}
\left\| F\left( \psi^*\gamma ^0 \psi\,,\,\psi^*\gamma ^0\gamma ^5 \psi \right)\psi (t) \right\|_{(H_{(s)}({\mathbb R}^3))^4}
\leq C_s\left\| \psi (t)\right\|_{(H_{(s)}({\mathbb R}^3))^4}\,.
\end{equation} 
For every $s \geq 6 $ the local Cauchy problem for (\ref{NDE_CP}) is well posed  in $C^0([0,T_s);(H_{(s)}({\mathbb R}^3))^4)  $ for some $0<T_s$. According to (\ref{B49}) and
\[
 \|\psi (t)\|_s   
 \leq 
C_s e^{\frac{1}{2} \delta_+ t}    + C_s  e^{\frac{1}{2} \delta_+ t} \int_0^t e^{-\frac{1}{2} \delta_+ \tau }  \|\psi (x,\tau )\|_s  \,  d\tau \,,\qquad t \in [0,T_s)\,,
\]
from the last inequality we conclude that $T_s=\infty$ and $\psi (t) \in (C_0^\infty({\mathbb R}^3))^4 $.  The equation (\ref{NDE_CP}) implies $\psi   \in C^1([0,\infty);(C_0^\infty({\mathbb R}^3))^4) $. Theorem is proved. \qed

\section{Large  time asymptotics  for large data solution}
\label{S4}
\setcounter{equation}{0}

\begin{theorem}
\label{T4.1A}
Let $\psi =\psi (x,t) $ be a solution of the problem   
\[ 
\cases{ \dsp\left(  i {\gamma }^0    \partial_0   +i e^{-Ht}\sum_{\ell=1,2,3}{\gamma }^\ell  \partial_\ell +i \frac{3}{2}    H {\gamma }^0     -m{\mathbb I}_4 +\gamma^0 V(x,t)\right)\psi   =F\left( \psi^* \gamma ^0 \psi\,,\,\psi^*\gamma ^0\gamma ^5 \psi \right)\psi  \,,\cr
\psi (x,0)=\Psi _0 (x)+\varepsilon \chi _0(x)\,,
}
\]
given by Theorem~\ref{T4.1}. Assume that  $F\left( \psi^* \gamma ^0 \psi\,,\,\psi^*\gamma ^0\gamma ^5 \psi \right)\psi$  is the Lipschitz continuous function with exponent $\alpha >0 $ in the space $H_{(6)}({\mathbb R}^3) $.
 
Then the limit
\begin{equation}
\label{Limit}
\lim_{t \to  \infty}   \int_0^{ t} S(0,\tau ) \gamma ^0F\left( \psi^* (x,\tau )  \gamma ^0 \psi(x,\tau )\,,\,\psi^* (x,\tau )\gamma ^0\gamma ^5 \psi (x,\tau )\right)\psi (x,\tau )\,d \tau   
\end{equation}
exists in the space  $H_{(6)}({\mathbb R}^3) $.
Furthermore, the solution $   \psi^+ (x,t)$  of the Cauchy problem for the free Dirac equation (\ref{DFree}),  
where
\begin{eqnarray*} 
\psi^+ _0 (x) 
& = & 
\Psi _0 (x)+\varepsilon \chi _0  (x)\\
&   &
-i\lim_{t \to  \infty}   \int_0^{ t} S(0,\tau )\gamma ^0 F\left( \psi^* (x,\tau )  \gamma ^0 \psi(x,\tau )\,,\,\psi^* (x,\tau )\gamma ^0\gamma ^5 \psi (x,\tau )\right)\psi (x,\tau )\,d \tau\,,
\end{eqnarray*}
 satisfies 
\[
\lim_{t \to +\infty}    \left\|  \psi (x,t)-    \psi^{+} (x,t) \right\|_{(H_{(6)}({\mathbb R}^3))^4}=0
\]
and  ${\mathscr{S}}^+\,:\,\psi_0  \longmapsto  \psi^{+}_0 $ is  the continuous    operator.

Moreover, if $V(x,t)=0$, then the function $\psi^+   (x,t) $ is given by (\ref{2.20}).
\end{theorem}
\medskip

\noindent
{\bf Proof.} It is enough to prove the convergence in the space $(H_{(6)}({\mathbb R}^3))^4$. 
The solution $ \psi $ can be written as in  (\ref{B33}), 
$ \psi =\Psi +\chi $, 
where the function $\Psi=\Psi (x,t) $ solves   (\ref{B31}) 
while the function $\chi =\chi  (x,t) $ solves   (\ref{B34}),
where
\[
|f_1\left(\chi \,,\,\Psi  \right)| =O\Big( |\chi |\big(|\chi |+|\Psi  |\big)^2\Big)\,.
\]
According to 
(\ref{B48}) 
\begin{eqnarray*}
\left\| F\left( \psi ^*\gamma ^0 \psi\,,\,\psi ^*\gamma ^0\gamma ^5\psi \right)\psi (t)\right\|_{(H_{(6)}({\mathbb R}^3))^4}
& \leq  &
C_s\left\| \psi(t) \right\|_{(H_{(6)}({\mathbb R}^3))^4}^{1+\alpha } \\
& \leq  &
C_se^{\frac{1}{2}\delta _+ (1+\alpha )t}\,.
\end{eqnarray*}
At the same time 
\begin{eqnarray*}
\left\| S(0,\tau )\gamma ^0 F\left(\psi ^*\gamma ^0 \psi\,,\,\psi ^*\gamma ^0\gamma ^5\psi \right)\psi (\tau )\right\|_{(H_{(6)}({\mathbb R}^3))^4}
& \leq  &
C_se^{ -\frac{1}{2}\delta _-  \tau }\left\| \psi(\tau ) \right\|_{(H_{(6)}({\mathbb R}^3))^4}^{1+\alpha }\\
& \leq  &
C_se^{ -\frac{1}{2}\delta _-  \tau }e^{\frac{1}{2}\delta _+ (1+\alpha ) \tau }
\end{eqnarray*}
that implies the convergence of (\ref{Limit}) in $(H_{(6)}({\mathbb R}^3))^4 $.
\qed

\medskip

\section{Nonexistence of global in time solution}
\label{S5}
\setcounter{equation}{0}

Consider the semilinear  Dirac equation 
\[
  \dsp 
\left(  i {\gamma }^0    \partial_0   +i e^{-Ht}\sum_{\ell=1,2,3} {\gamma }^\ell  \partial_\ell +i \frac{3}{2}    H {\gamma }^0     -m{\mathbb I}_4+i\gamma^0V(x, t) \right)\psi=iG(\psi ){\gamma }^0\psi  \,,
\]
where $H \in {\mathbb R}$, $m \in \C$, $V^*(x, t)=V(x, t)$, and the matrix-valued term  $ G(\psi )$ commutes with ${\gamma }^0 $, $
{\gamma }^0 G(\psi )=G(\psi ){\gamma }^0 $. 
The equation can  also be written in the equivalent form of the following symmetric hyperbolic system 
\[
  \dsp 
{\mathscr{D}}_{dS}(x,t, \partial)\psi=G(\psi ) \psi  \,.
\]
\begin{lemma}
\label{L1}
Let $H \in {\mathbb R}$, $m \in \C$. Then for the derivative of the energy integral  we have
\[
\frac{d}{d t} E(t)
=
  \int_{{\mathbb R}^3}\left(  2\Re ( G_{jk}(\psi )   \psi_k (x,t) \ol{\psi_j (x,t)}) - 3    H |\psi (x,t)|^2   + 2(\Im (m)) \psi^* (x,t)  {\gamma }^0  \psi (x,t)  \right) dx.
\]
\end{lemma} 
\medskip

\ndt
{\bf Proof.} The arguments have been used in the proof of Lemma~\ref{L1.2}  complete  the proof of  lemma. 
 \qed 
\medskip

\subsection{Nonexistence of global  solution in the expanding universe}

The next theorem gives blow up result for the solution with the large data.
\begin{theorem}
\label{TBUH+}
Consider the Cauchy problem 
\begin{equation}
\label{CPHP}
\cases{
  \dsp 
\left(  i {\gamma }^0    \partial_0   +i e^{-Ht}\sum_{\ell=1,2,3}{\gamma }^\ell  \partial_\ell+i \frac{3}{2}    H {\gamma }^0     -m{\mathbb I}_4 +i\gamma^0V(x, t)\right)\psi=iG(\psi ){\gamma }^0\psi  \,,\cr
\psi (x,0) =\psi_0 (x )
}
\end{equation}
with $ \psi_0 (x )$ such that $
\mbox{\rm supp\,} \psi_0 (x ) \subseteq  \{x \in {\mathbb R}^3\,|\,|x| \leq R\}$. 
Assume that $H>0$, $V^*(x, t)=V(x, t)$,  and 
\begin{eqnarray}
&  &
 G(\zeta ) =O(|\zeta |),  \quad  {\gamma }^0 G(\zeta )=G(\zeta ){\gamma }^0 , \hspace*{0.5cm}  \Re (G(\zeta )\zeta  ,\ol{\zeta } )\geq c_0|\zeta |^{2+\alpha },\quad \alpha >0, \nonumber \\
\label{1.6}
&  &
\int_{{\mathbb R}^3}|\psi_0 (x )|^{2} \,dx    > \left( \frac{3H+2|\Im (m)|  }{c_0 } \right) ^{  2 /\alpha} \left( R +\frac{1}{H}  \right)^{3} \,.
\end{eqnarray}
Then the solution $\psi $ of (\ref{CPHP}) that obeys the finite propagation speed property 
$\mbox{\rm supp\,} \psi (x,t) \subseteq 
 \big\{x \in {\mathbb R}^3\,\big|  \,|x| \leq   $ $R +\phi (t) \big\}$ 
 blows up at finite time. More precisely, for the time $T$ defined by
 \[
T = - \frac {2}{\alpha (3H+2|\Im (m)|)} \ln \left( 1-\frac {3H+2|\Im (m)|}{c_0}   \left( \int_{{\mathbb R}^3}|\psi_0 (x  )|^{2} \,dx   \right)^{- \alpha /2} \left( R +\frac{1}{H}  \right)^{ 3\alpha /2} \right)
 \]
the following is true 
 \[\lim_{t \nearrow T}\int_{{\mathbb R}^3}|\psi (x,t )|^{2} \,dx =\infty\,.
 \]
\end{theorem}
\medskip

\medskip

\ndt
{\bf Proof.} According to 
Lemma~\ref{L1} 
if  $H \in {\mathbb R}$, $m \in \C$, then
\begin{eqnarray*}
\frac{d}{d t} E(t)
& = &
  \int_{{\mathbb R}^3}\left(  2\Re ( G_{jk}(\psi )   \psi_k (x,t) \ol{\psi_j (x,t)}) - 3    H |\psi (x,t)|^2   + 2\Im (m) \psi (x,t)  {\gamma }^0 \ol{\psi (x,t)  }  \right) \,dx\\
& \geq  &
  \int_{{\mathbb R}^3}\left( c_0 |\psi (x,t)|^{2+\alpha } - 3    H |\psi (x,t)|^2   + 2\Im (m) \psi (x,t)  {\gamma }^0 \ol{\psi (x,t)  }  \right) \,dx\\
& \geq  &
c_0  \int_{{\mathbb R}^3}|\psi (x,t)|^{2+\alpha } \,dx-(3H+2|\Im (m)| )\int_{{\mathbb R}^3} |\psi (x,t)|^2  \,dx\,.
\end{eqnarray*}
Further, if the solution obeys the finite speed propagation property, we obtain
\[
\int_{{\mathbb R}^3}  |\psi (x,t)|^2  \,dx
 \leq 
\left( \int_{\mbox{\rm supp\,} \psi}   |\psi (x,t)|^{2+\alpha }  \,dx \right)^{2/(2+\alpha )}\left( R +\phi (t)   \right)^{3\alpha /(2+\alpha )}  \,.
\]
It follows
\begin{equation}  
\label{5.4b}
\left( R +\phi (t)  \right)^{-3\alpha /2} \left( \int_{{\mathbb R}^3}  |\psi (x,t)|^2  \,dx\right)^{(2+\alpha )/2}
 \leq  
\left( \int_{\mbox{\rm supp\,} \psi}   |\psi (x,t)|^{2+\alpha }  \,dx \right)\,.
\end{equation}
Then
\[
\frac{d}{d t} E(t)
\geq  
K(t) E(t)^{(2+\alpha )/2} 
-AE(t)\,,
\]
where
\[
K(t)
 :=  
c_0  \left( R +\phi (t)  \right)^{-3\alpha /2}\,, \quad
A   :=  
 (3H+2| \Im (m) | ) \,.
\] 
Hence
\[
\frac{d}{d t} \left( E(t)e^{At} \right) 
\geq  
K(t)   e^{-A\alpha t  /2}     \left( E(t)e^{At} \right)^{(2+\alpha )/2}\,.  
\]
For the function  
\[
 F(t)
  :=   
  E(t)e^{At}    
\]
the inequality leads to 
\[
 \frac{d}{d t} F(t) ^{ -\frac{\alpha}{2} } 
  \leq   
-\frac{\alpha}{2} K(t)   e^{-A\frac{\alpha}{2}  t  }   \,.
\]
After integration we obtain
\[
   F(t) ^{ -\frac{\alpha}{2}}   
  \leq  
F(0) ^{ -\frac{\alpha}{2} } -\frac{\alpha}{2}  c_0\int_0^t   \left( R +\phi (s)  \right)^{-3\frac{\alpha}{2}}    e^{-A  s \frac{\alpha}{2} } \,ds  \,.   
\]
Since the function $K(s)$ is monotonically decreasing, we obtain
\[
   F(t) ^{ -\frac{\alpha}{2}}   
 \leq   
F(0) ^{ -\alpha /2} -\frac{1}{A}  c_0\left( R +\frac{1}{H}  \right)^{-3\frac{\alpha}{2}}(1- e^{-A  t \frac{\alpha}{2} })  \,.      
\]
The condition (\ref{1.6}) guarantees that the solution blows up no later than  time $T$ such that
\[
F(0) ^{ -\alpha /2} = \frac{1}{A }  c_0\left( R +\frac{1}{H}  \right)^{-3\frac{\alpha}{2}}(1- e^{-A T\frac{\alpha}{2} })
\]
provided that $F(0)$ is sufficiently large. 
Theorem is proved. \qed
\medskip

\subsection{Nonexistence of global  solution in the contracting universe}

In the next theorem  $ F\left(a,b;c ;z\right) $ is the hypergeometric function (see, e.g., \cite{B-E}).  
\begin{theorem}
\label{T5.3}
Consider the Cauchy problem (\ref{CPHP}) 
with $ \psi_0 (x )$ such that $
\mbox{\rm supp\,} \psi_0 (x ) \subseteq  \{x \in {\mathbb R}^3\,|\,|x| \leq R\}$.  
Assume that $H<0$, $V^*(x, t)=V(x, t)$, and for $G$, $m$, and  $ H $  there is a constant $c_{G,H,m}>0$ such that   
\begin{eqnarray}
 \label{1.5-}
&  &
G(\zeta ) =O(|\zeta |), \quad  {\gamma }^0 G(\zeta )=G(\zeta ){\gamma }^0  \quad \mbox{\rm  for all} \quad  \zeta  \in \C^4,\\
 \label{1.6-}
&  &  
2 \Re (G(\zeta )\zeta  , \ol{\zeta}  )   - 3    H|\zeta |^2 + 2\Im (m)(\zeta ,\gamma ^0 \ol{\zeta})   \geq c_{G,H,m}|\zeta |^3 \quad \mbox{\rm  for all} \quad  \zeta  \in \C^4  \,.
\end{eqnarray}
Then the solution $\psi $ of (\ref{CPHP}) that obeys the finite propagation speed property
$\mbox{\rm supp\,} \psi (x,t) \subseteq 
\big\{x \in {\mathbb R}^3\,\big|  \,|x| \leq  $ $  R +\phi (t) \big\}$ 
 blows up at finite time. More precisely, there  is   $T_{ls}<\infty$ defined by 
 \[
\int_{{\mathbb R}^3}|\psi_0 (x  )|^{2} \,dx
  =  
 \left\{ \frac{c_{G,H,m}\alpha }{2}\int_0^{T_{ls}}\left( R +\phi (s)  \right)^{-3\alpha /2}  \,ds\right\}^{ -2/\alpha }\,,
\]
depending on  $c_0$, $\alpha $, $m$, $\psi _0$, and $H$  such that if  
\[
\int_{{\mathbb R}^3}|\psi_0 (x  )|^{2} \,dx
  >  
 \left\{ \frac{ c_{G,H,m}R^{1-\frac{3 }{2}}}{3  } F\left(1,1;\frac{3 \alpha }{2}+1;H R+1\right)\right\}^{ 2/\alpha }\,,
\]
then
 \[
\lim_{t \nearrow T_{ls}}\int_{{\mathbb R}^3}|\psi (x,t )|^{2} \,dx =\infty\,.
 \]
\end{theorem}
\medskip

\ndt
{\bf Proof.}
From Lemma~\ref{L1}
\[
\frac{d}{d t} E(t)
=
  \int_{{\mathbb R}^3}\left(  2\Re ( G_{jk}(\psi )   \psi_k (x,t) \ol{\psi_j (x,t)}) - 3    H |\psi (x,t)|^2   + 2(\Im (m)) \psi (x,t)  {\gamma }^0 \ol{\psi (x,t)  }  \right) dx
\]
 and conditions (\ref{1.5-}), (\ref{1.6-})  we derive 
\[
\frac{d}{d t} E(t)
  \geq  
 c \int_{{\mathbb R}^3}  |\psi (x,t)|^{2+\alpha }  \,dx\,.
\]
Then,  since the solution obeys the finite propagation speed property, we obtain (\ref{5.4b}) 
and, consequently, 
\[
E(t)^{-(2+\alpha )/2}\frac{d}{d t} E(t)
  \geq  
c_{G,H,m} \left( R +\phi (t)  \right)^{-3\alpha /2} \,.
 \]
 We integrate the last inequality and obtain
\[
-E(t)^{-\alpha /2} +E(0)^{-\alpha /2}
  \geq  
\frac{\alpha }{2}c_{G,H,m} \int_0^t\left( R +\phi (t)  \right)^{-3\alpha /2} \,ds 
\]
that leads to
\[
E(t)
  \geq  
\left(\frac{1}{E(0)^{ \alpha /2}}- \frac{c_{G,H,m}\alpha }{2} \int_0^t\left( R +\phi (t)   \right)^{-3\alpha /2}\,ds\right)^{-2/\alpha }.
\]
Now for $H<0$ we calculate
\begin{eqnarray*}
&    &
\int_0^t\left( R + \phi (s)  \right)^{-3\alpha /2}  \,ds\\
&= &
\frac{2}{3 \alpha }\left\{R^{1-\frac{3 \alpha }{2}} F\left(1,1;\frac{3 \alpha }{2}+1;H R+1\right)\right.\\
&  &
\left.+|H|^{\frac{3 \alpha }{2}-1}\left((H R+1) e^{H t}-1\right) \left(|H| R+e^{-H t}-1\right)^{-\frac{3 \alpha }{2}} F\left(1,1;\frac{3 \alpha }{2}+1;e^{H t} (H R+1)\right)\right\}\,.
\end{eqnarray*}
Here
\[
\lim_{t \to \infty} \int_0^t\left( R +\phi (s)  \right)^{-3\alpha /2}  \,ds
  =  
\frac{2 }{3 \alpha }R^{1-\frac{3 \alpha }{2}} F\left(1,1;1+\frac{3 \alpha }{2};H R+1\right),\quad H<0,
\]
that gives a blowup of the solution  with the large data such that 
\[
\frac{1}{E(0)^{ \alpha /2}}
  <  
\frac{c_{G,H,m}\alpha }{2} \frac{2 }{3 \alpha } R^{1-\frac{3 \alpha }{2}}F\left(1,1;1+\frac{3 \alpha }{2};H R+1\right),\quad H<0\,,
\]
and with the lifespan $T_{ls} $ that can be obtained from 
\[
\frac{1}{E(0)^{ \alpha /2}}
  =  
\frac{c_{G,H,m}\alpha }{2}\int_0^{T_{ls}}\left( R +\phi (s)  \right)^{-3\alpha /2}  \,ds\,.
\]
The theorem is proved. \qed

\begin{remark}
We do not know whether  
for small data $E(0) $ such that 
\[
\int_{{\mathbb R}^3}|\psi_0 (x  )|^{2} \,dx
<  
 \left\{ \frac{ c_{G,H,m}R^{1-\frac{3 }{2}}}{3  } F\left(1,1;\frac{3 \alpha }{2}+1;H R+1\right)\right\}^{ 2/\alpha }\,,
\]
the global solution exists.  Finally, we note that if $ \psi_0 (x ) \in C_0^\infty(\R^3) $, then   the solution of the problem obeys the finite propagation speed property, which  is proved in Section~\ref{SS5.4}.
\end{remark}

\section{Finite speed propagation property}
\label{SS5.4}

We are going to prove that the dependence domain for the classical solution $u(t,x)$ at the point $(T,x_0)$  of the semilinear equation coincides with the dependence domain of the solution of the linear equation (\ref{LDE}). 
More precisely, for  $(T,x_0) \in[0,\infty)\times \R^3$ with  $T>0$  let  
\[
\Sigma ^- {(T,x_0)}:= \left\{(t,x) \in [0,T]\times \R^3\,\Big|\, |x-x_0|=-(\phi (t) -\phi (T)) \right\}
\]
be  a part of the backward   ``curved light cone'' (nullcone),  where  $\phi (t):= (1-e^{-Ht} )/H$. Let  also     
\[
D_- {(T,x_0)}= \left\{(t,x) \in [0,T]\times \R^3\,\Big|\, |x-x_0|\leq -\left(\phi (t) -\phi (T)  \right)\right\}
\]
be  the region defined in (\ref{1.3D}), whose       boundary contains   $\Sigma ^- {(T,x_0)} $. 
In the proof of the next theorem, we  follow \cite{John}. (See also \cite[Ch.2,  \textsection 6]{Taylor_I} and \cite[Ch.16,  \textsection 1]{Taylor_III}.)
\begin{theorem}
Let $\psi $ be a $C^1$ solution of the equation 
\begin{equation}
\label{5.4}
  \dsp 
\left(      \partial_0   + e^{-Ht}\sum_{\ell=1,2,3}\alpha ^\ell  \partial_\ell +\frac{3}{2}    H  {\mathbb I}_4    +im{\gamma }^0 -iV (x,t) \right)\psi=F(\psi ,\psi ')  \,,
\end{equation}  
in the backward curved light cone $D_- (T,x_0)$ through $(T,x_0) \in (0,\infty)\times \R^3$ with  $T>0$. Assume that  the   potential $  V \in C([0,\infty) \times  \R^3 )  $  and   the nonlinear term $F(\psi ,\psi ') \in C^1 $ is such that
\begin{equation}
\label{5.F}
F(0,\psi ')=0 \quad \mbox{ for all}\quad \psi '=(\partial_t \psi , \nabla_x \psi ) \,.
\end{equation} 
If 
\begin{equation}
\label{5.IC}
\psi (x,0)=0 \quad \mbox{for all}\quad x \in D_- (T,x_0)\cap \{ t=0\}\,,
\end{equation}
then $\psi $ vanishes in $ D_- (T,x_0)$.
\end{theorem}
\medskip

\ndt
{\bf Proof.}  
The interior of the domain $D_- {(T,x_0)} $ can be filled up by one-parameter family  of the smooth spacelike surfaces $\Sigma _s^- {(T,x_0)}$, where $s$ is a parameter. In order to find the equation $t=\tau  (s,x)$ of such surfaces we slightly modify  the equation of the light cone 
$
\phi (T)  - \phi (  t) = |x-x_0|$, $\, t \leq T 
$,  
by equipping it with the parameter $s$  
in the following way
\[
  (\phi (T)  - \phi (  t)  )^2 =(s- \phi ( T) )^2  +(\phi (T))^{-2} \left(2 s \phi (T) -s^2\right)|x-x_0|^2, \quad 0\leq s\leq  \phi (T)\,.
\] 
Then for the values of the parameter $ 0\leq s<  \phi (T)$, by the implicit function theorem,  the last equation can be solved for $t$ since $\phi' (  t) (\phi (T)  - \phi (  t)  )\not= 0$. The solution   $t=\tau (s,x)$  is 
 \[
\tau(s,x)=-\frac{1}{H}  \ln \Big( 1- H\phi (T) +H[(s- \phi ( T) )^2  +(\phi (T))^{-2} \left(2 s \phi (T) -s^2\right)|x-x_0|^2]^{1/2}\Big)\,.  
\]
Here
 \[
\tau(0,x)=0, \quad \lim_{s \to \phi (T) } \tau(s,x) 
=-\frac{1}{H}  \ln \Big( 1-H(\phi ( T)-|x-x_0|)\Big)
\,.  
\]
For every given $s$, $ 0\leq s\leq  \phi (T)$,  we  consider $x$ such that 
\[
|x-x_0|^2\leq \frac{ (\phi (T))^{ 2}[  (\phi (T)  - \phi (  t)  )^2 -(s- \phi ( T) )^2 ]}{\left(2 s \phi (T) -s^2\right)} \,.
\]
The region  bounded by the surface  $\Sigma _s^- {(T,x_0)}$ is 
\[
D_{-,s} {(T,x_0)} =\left\{(t,x) \,\Big|\, 0\leq t\leq \phi (s,x),   |x-x_0|^2\leq \frac{ (\phi (T))^{ 2}[  (\phi (T)  - \phi (  t)  )^2 -(s- \phi ( T) )^2 ]}{\left(2 s \phi (T) -s^2\right)}  \right\}\,.
\]
Hence 
\[
D_- {(T,x_0)}= \bigcup_{  0\leq s\leq  \phi (T)} D_{-,s} {(T,x_0)}\,.
\]  
The surface $\Sigma _s^- {(T,x_0)}$ is space-like. Indeed, its outward unit normal at $(\tau  (s,x),x)$   is 
\[
n(s,x)= \frac{(1, - \nabla_x \tau  (s,x))}{\sqrt{1+|\nabla_x \tau  (s,x)|^2}}\,,
\]
where
\[
|\nabla_x \tau  (s,x)|
  =  
\frac{e^{H\tau  (s,x)}(\phi (T))^{-2} \left( 2s \phi (T) -s^2\right)|  x -x_0 |}{[(s- \phi ( T) )^2  +(\phi (T))^{-2} \left(2 s \phi (T) -s^2\right)|x-x_0|^2]^{1/2}}\,,
\]
and, consequently, 
\[
n(s,x)^*g^{-1}(\tau  (s,x))n(s,x)= (1 -e^{-2H\tau  (s,x)}|\nabla_x \tau  (s,x)|^2)\frac{1}{ 1+|\nabla_x \tau  (s,x)|^2 }  >0\,.
\]
Next,  we apply the energy method  to the equation (\ref{5.4}). 
We write the identity
\begin{eqnarray*}
&  &
  \dsp 
   \partial_0 |\psi|^2    +\sum_{\ell=1,2,3} \partial_\ell ( e^{-Ht}\psi ^*  \alpha ^\ell  \psi) +3   H |\psi|^2    -2\Im (m)\psi ^*  {\gamma }^0 \psi+2\psi ^*  \Im (V(x,t))  \psi \\
& = &
\psi ^* F(\psi ,\psi ') + F(\psi ,\psi ')^* \psi \,.  
\end{eqnarray*}  
By the divergence theorem and the vanishing initial conditions (\ref{5.IC}) we obtain
\begin{eqnarray*}
 &  &
\int_{D_{-,s} {(T,x_0)}}  \dsp 
  \Bigg(  \partial_0 |\psi|^2    +\sum_{\ell=1,2,3} \partial_\ell ( e^{-Ht}\psi ^*  \alpha ^\ell  \psi) \Bigg) dt \,dx \\
& =  &
\int_{\Sigma _s^- {(T,x_0)}}  \dsp 
  \left(  |\psi|^2    +\sum_{\ell=1,2,3}  e^{-Ht}\psi ^*  \alpha ^\ell  \psi (\partial_\ell \tau  (s,x)  ) \right) \frac{1}{\sqrt{1+|\nabla_x \tau  (s,x)|^2}}\,d \sigma \,.
 \end{eqnarray*}  
  Hence, 
\begin{eqnarray}
\label{5.5}
& &
\int_{\Sigma _s^- {(T,x_0)}}  \dsp 
  \left(  |\psi|^2    +\sum_{\ell=1,2,3}  e^{-Ht}\psi ^*  \alpha ^\ell  \psi (\partial_\ell \tau  (s,x)  ) \right) \frac{1}{\sqrt{1+|\nabla_x \tau  (s,x)|^2}}\,d \sigma \nonumber \\
&  &
 +\int_{D_{-,s} {(T,x_0)}}  \dsp 
  \left( 3   H |\psi|^2    -2\Im (m)\psi ^*  {\gamma }^0 \psi +2\psi ^*  \Im (V(t))  \psi\right) dt \,dx  \nonumber \\
& = &
\int_{D_{-,s} {(T,x_0)}} \left(  \psi ^* F(\psi ,\psi ') + F(\psi ,\psi ')^* \psi  \right) dt \,dx\,.  
\end{eqnarray}
We are going to estimate
\[
\sum_{\ell=1,2,3}  e^{-H\tau  (s,x) }\psi ^*  \alpha ^\ell  \psi (\partial_\ell \tau  (s,x)  )  \frac{1}{\sqrt{1+|\nabla_x \tau  (s,x)|^2}}
\]
 on the surface $\Sigma _s^- {(T,x_0)}$. If we restrict the parameter $s$ by $0\leq s\leq s_0< \phi (T)$, then
\begin{eqnarray*}
e^{-H\tau  (s,x)}|\nabla_x \tau  (s,x)|
& = &
\frac{(\phi (T))^{-2} \left( 2s \phi (T) -s^2\right)|x-x_0|}{[(s- \phi ( T) )^2  +(\phi (T))^{-2} \left(2 s \phi (T) -s^2\right)|x-x_0|^2]^{1/2}}\\
& \leq &
\vartheta (s_0)<1\,.
\end{eqnarray*}
Consider the hermitian matrix 
\[
A={\mathbb I}_4- \sum_{\ell=1,2,3}    \alpha ^\ell  a_\ell,\quad a_\ell:= e^{-H\tau  (s,x)}\frac{\partial_\ell \tau  (s,x) }{\sqrt{1+|\nabla_x \tau  (s,x)|^2}},\quad \ell=1,2,3.
\]
It has two double  eigenvalues  
\[
1-\sqrt{a_1^2+a_2^2+a_3^2}>0\quad  \mbox{\rm and}\quad 1+\sqrt{a_1^2+a_2^2+a_3^2}>0  \quad  \mbox{\rm for all}\quad s \in [0,s_0]. 
\]
Hence, there is $\delta (s_0)<1 $ such that 
\[
\left| \sum_{\ell=1,2,3}  e^{-H\tau  (s,x) }\psi ^*  \alpha ^\ell  \psi (\partial_\ell \tau  (s,x)  )  \frac{1}{\sqrt{1+|\nabla_x \tau  (s,x)|^2}} \right|\leq 
\delta (s_0) | \psi |^2 \quad  \mbox{\rm for all}\quad s \in [0,s_0]\,.
\]
The equation (\ref{5.5}) and condition (\ref{5.F}) imply
\begin{eqnarray}
\label{5.5b}
&  &
\int_{\Sigma _s^- {(T,x_0)}}  \dsp 
   |\psi|^2   \frac{1}{\sqrt{1+|\nabla_x \tau  (s,x)|^2}}d \sigma \nonumber  \\
& \leq  &
C(s_0) \int_0^s \int_{\Sigma _\lambda ^- {(T,x_0)}}  \dsp 
 |\psi|^2    \frac{\partial_\lambda \tau   (\lambda ,x)}{\sqrt{1+|\nabla_x \tau   (\lambda ,x)|^2}}\,d \sigma\,  d\lambda  
\quad  \mbox{\rm for all}\quad s \leq s_0\,.
\end{eqnarray}
Here
\[
\partial_s \tau   (s ,x)= \frac{ \phi ( T)-s   -  s(\phi (T))^{-2} ( \phi ( T)-1 ) |x-x_0|^2}{\phi' ( \tau   (s ,x))  (\phi (T)  - \phi ( \tau   (s ,x)) }\,, \quad 0\leq s<  \phi (T)\,.
\] 
If we set
\[
I(s)=\int_{\Sigma _s^- {(T,x_0)}}  \dsp 
|\psi|^2   \frac{1}{\sqrt{1+|\nabla_x \tau |^2}}\,d \sigma \,,
\]
then from (\ref{5.5b}) it follows
\[
I(s)
\leq 
C(s_0) \left(\max_{0\leq t\leq s_0 }\left|\partial_\lambda \tau   (t ,x) \right|\right) \int_0^s  I(\lambda ) \, d\lambda  
\quad  \mbox{\rm for all}\quad s \leq s_0\,,
\]  
and 
Gronwall's inequality completes the proof of theorem.
\qed

\begin{small}

\end{small}
\end{document}